\documentclass[12pt]{article}
\usepackage[dvips]{color}
\usepackage{epsfig}
\usepackage{amsmath}
\usepackage{graphicx}

\begin{document}

\title{\bf Phase Transition of charged Rotational
Black Hole and Quintessence}

\author{{Kh. Jafarzade $^{a}$ \thanks{Email: kh.Jafarzadeh@stu.umz.ac.ir} \hspace{1mm},
J. Sadeghi $^{a}$\thanks{Email:
pouriya@ipm.ir}} \\
$^a${\small {\em Sciences Faculty, Department of Physics, University
of Mazandaran,}}\\{\small {\em P. O. Box 47415-416, Babolsar,
Iran}}}

\maketitle
\begin{abstract}
In this paper, we calculate thermodynamical quantity of
Kerr-Newman-AdS black hole solution in quintessence matter. Then, we
show that how the rotation and cosmological parameters effect to the
thermodynamics properties of black hole. Also, we investigate both
types of phase transition for different values of $\omega$ parameter
in extended phase space.  We notice that type one of phase
transition occurs for $P<0.42$ and $a<0.5$. And also we see that the
phase transition point shifts to higher entropy when pressure $P$,
rotation parameter $a$ and $\alpha$ increase. Also, we find that by
changing parameter $\omega$ from -1 to $-\frac{1}{3}$, the critical
point shifts to higher entropy. Then we study type two of phase
transition and show critical points increase by increasing parameter
$\alpha$. Also, we show that the critical point shifts to higher
entropy when $\alpha$, $\omega$ and rotation parameter $a$ decrease.
Finally, we find that by decreasing pressure the first critical
point shifts to lower entropy and second critical point shifts to
higher entropy.

\end{abstract}
\section{Introduction}
As we know black holes play an important role in physics especially
quantum gravity. One of the interesting methods  to study the
quantum gravity is black hole thermodynamics in AdS spacetime [1,2].
First time Hawking and Bekenstein investigated  the black hole
thermodynamic. They found that all laws of black hole mechanics are
similar to ordinary thermodynamics where surface gravity,  mass  and
area of black hole are related to the temperature, energy and
entropy respectively [3]. The study  of black hole thermodynamic
will be interesting  with presence of a negative cosmological
constant. Because such a cosmological constant play important role
in holography and $AdS/CFT$. From AdS/CFT point of view,
asymptotically AdS black hole spacetimes admit a gauge duality
description with dual thermal field theory. Such theory lead us to
interesting phenomenon which is called phase transition. For the
first time, the phase transition was studied by Hawking and Page
[4]. They found a phase transition between Schwarzschild-AdS black
holes and thermal radiation which is known as Hawking-Page phase
transition. After that, a lot of studies were done in context of AdS
black hole thermodynamics especially  in extended phase space [5-9].
In AdS black hole thermodynamics, the cosmological constant of the
spacetime treat as pressure and its conjugate quantity as a
thermodynamic volume. But in extended phase space,  it appear as a
thermodynamical variable in the first law of black hole
thermodynamics. Here, remarkable matter is that in extended phase
space the mass of black hole $M$ is not related to energy but it is
associated with enthalpy  by $M=H\equiv U+PV$ [3], so the first Law
is expressed as,
\begin{equation}
dM=TdS+VdP+\Phi dQ+\Omega dJ.
\end{equation}
The phase transition plays a key role to study thermodynamical
properties of a system near critical point [10-14]. One of the
approaches to investigate phase transition is studying the behavior
of the heat capacity in different ensembles. In the study of heat
capacity, we encounter two types of phase transition as type one and
type two. The type one is related to changes of signature in heat
capacity where roots of heat capacity are representing phase
transitions. The type two is related to divergency of the heat
capacity. It means that the singular points of the heat capacity are
representing the phase transitions [15,16]. Also, the heat capacity
is an interesting thermodynamical quantity to determine stability
and instability of the black hole. For general black hole heat
capacity is always negative which shows such a black hole is
unstable and produces Hawking radiation. But with presence of charge
and rotation parameters of black hole the heat capacity could be
positive and the phase transition will happen. But the study of
phase transition will be interesting when black holes are surrounded
by quintessence or other matters. In 2003, Kiselev obtained the
Einstein equation's solution for the quintessence matter around
schwarzschild black hole [17]. These solution is written in terms of
$\omega$ and $\alpha$. The state parameter $\omega$ is defined by
the equation of state $p=\omega\rho$ where $p$ and $\rho$ are the
pressure and energy density of the quintessence respectively.
Quintessence is a dark energy model with the state parameter
$-1<\omega<1$ which the case of $\omega = \frac{1}{3}$ represents
radiation and case of $\omega = 0$ represents dust around black hole
[18]. To describe the late-time cosmic acceleration, the $\omega$ is
restricted to $-1<\omega<-\frac{1}{3}$ but $\omega$ will not equal
$0, -1, -\frac{1}{3}$ [19-23]. The parameter $\omega$ is an
important parameter to determine the property of spacetime metric.
The spacetime has the asymptotically flat solution for
$(-\frac{1}{3} < \omega< 0)$ and it has de Sitter
horizon for $(-1< \omega< -\frac{1}{3})$.\\
The rotation Kiselev and Kerr-Newman kiselev solution was determined
in Ref. [24-27] and also Kerr-Newman-AdS solution in the
quintessence was obtained in [28]. Then some efforts were made in
context of thermodynamic and phase transition  for Kiselev and Kerr
Kiselev black hole [18,29]. In this paper we want to study
thermodynamical relation and phase transition for Kerr-Newman-AdS
solution in the quintessence.\\
The outline of the paper is as follows. In section II, we review
Kerr-Newman-AdS black hole solution in quintessence matter and
consider behavior of temperature. In section III, we study the phase
transition of this black hole for different value of parameter
$\omega$. Finally, in section IV we have some results and
conclusion.

\section{ Kerr-Newman-AdS black hole solution with quintessential energy}
The Kerr-Newman AdS metric in quintessence matter is expressed as
follows [28],
\begin{equation}
ds^{2}=\frac{\Sigma^{2}}{\Delta_{r}}dr^{2}+\frac{\Sigma^{2}}{\Delta_{\theta}}d\theta^{2}+\frac{\Delta_{\theta}sin^{2}\theta}{\Sigma^{2}}(a\frac{dt}{\Xi}-(r^{2}-a^{2})\frac{d\phi}{\Xi})^{2}-\frac{\Delta_{r}}{\Sigma^{2}}(\frac{dt}{\Xi}-a\sin^{2}\theta
\frac{d\phi}{\Xi})^{2},
\end{equation}

where
\begin{eqnarray}
\Delta_{r}=r^{2}-2Mr+a^{2}+Q^{2}+\frac{r^{2}}{\ell^{2}}(r^{2}+a^{2})-\alpha
r^{1-3\omega},\nonumber
\\&& \hspace{-98mm}
\Delta_{\theta}=1-\frac{a^{2}}{\ell^{2}}\cos^{2}\theta~~~~~~~~
\Xi=1-\frac{a^{2}}{\ell^{2}}
~~~~~~~~\Sigma^{2}=r^{2}+a^{2}\cos^{2}\theta.
\end{eqnarray}
The cosmological constant is $\Lambda=-\frac{3}{\ell^{2}}$ which
interpret as a thermodynamic pressure $P$ by,
\begin{equation}
P=-\frac{\Lambda}{8\pi}=\frac{3}{8\pi \ell^{2}}.
\end{equation}
The mass of black hole is determined by $\Delta_{r}(r_{+})=0$, so
\begin{equation}
M=\frac{(r_{+}^{2}+a^{2})(r_{+}^{2}+\ell^{2})+Q^{2}\ell^{2}-\alpha
\ell^{2}r^{1-3\omega}}{2r_{+}\ell^{2}},
\end{equation}
the area of event horizon of the black hole is given by,
\begin{equation}
A=4\pi\frac{(r_{+}^{2}+a^{2})}{\Xi}.
\end{equation}
The Bakenstein-Hawking entropy is given by
\begin{equation}
S=\frac{A}{4}=\pi\frac{(r_{+}^{2}+a^{2})}{\Xi},
\end{equation}
and Hawking temperature which is related to surface gravity is
defined by,
\begin{equation}
T_{H}=\frac{\kappa(r_{+})}{2\pi}=\lim_{\theta=0, r\rightarrow
r_{+}}\frac{\partial_{r}\sqrt{g_{tt}}}{2\pi\sqrt{g_{rr}}},
\end{equation}
Thus, the thermal temperature of Kerr-Newman-AdS solution in
quintessential dark energy is obtained as,
\begin{equation}
T=\frac{r_{+}}{4\pi\Xi(r_{+}^{2}+a^{2})}\bigg(\frac{3r_{+}^{2}}{\ell^{2}}+\frac{a^{2}}{\ell^{2}}+1-\frac{a^{2}+Q^{2}}{r_{+}^{2}}+3\alpha\omega
r^{-1-3\omega}\bigg).
\end{equation}
One can rewrite the temperature in terms of $S$ and $P$ which is
given by,
\begin{equation}
T=\frac{\sqrt{\frac{S\Xi}{\pi}-a^{2}}}{4S\Xi^{2}}\bigg(\frac{16\pi
P}{3}(\frac{S\Xi}{\pi}-a^{2})+\frac{8SP\Xi}{3}+1-\frac{(a^{2}+Q^{2})}{\frac{S\Xi}{\pi}-a^{2}}+3\alpha\omega(\frac{S\Xi}{\pi}-a^{2})^{\frac{-1-3\omega}{2}}\bigg).
\end{equation}
We plot Figure (1) and (2) to study the behavior of the temperature.
As we see in figure (1), rotation parameter $a$ and pressure $P$
play an important role in the temperature. The positive range of
temperature (physical state) increases by decreasing rotation and
pressure. And also, the temperature is always negative for $P>0.42$
and $a>0.5$ which Represents a unphysical state. Figure (2) shows
that $\omega$ and $\alpha$ parameters have small influence on sign
of temperature.
\begin{figure}
\hspace*{1cm}
\begin{center}
\epsfig{file=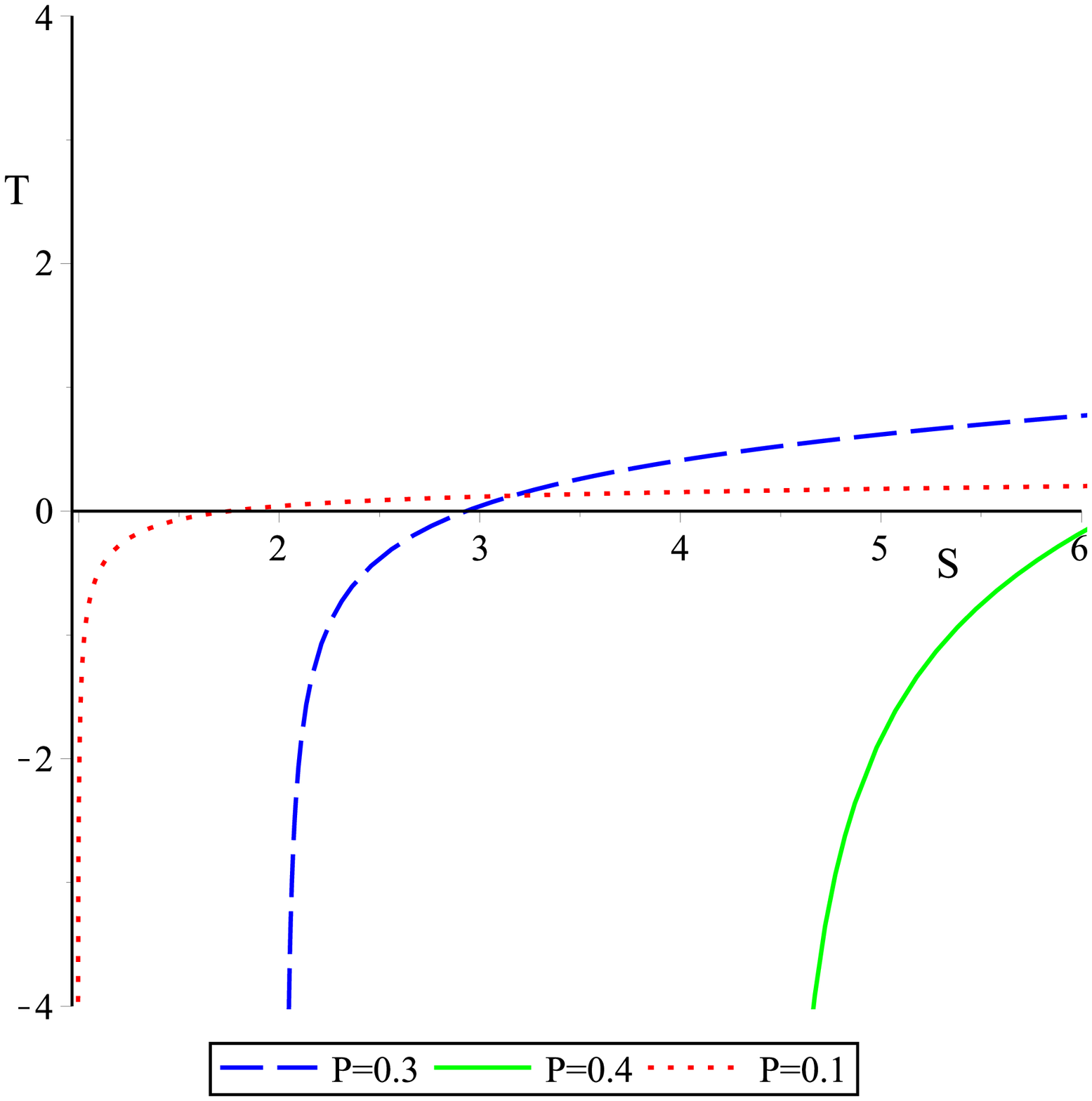,width=7cm}
\epsfig{file=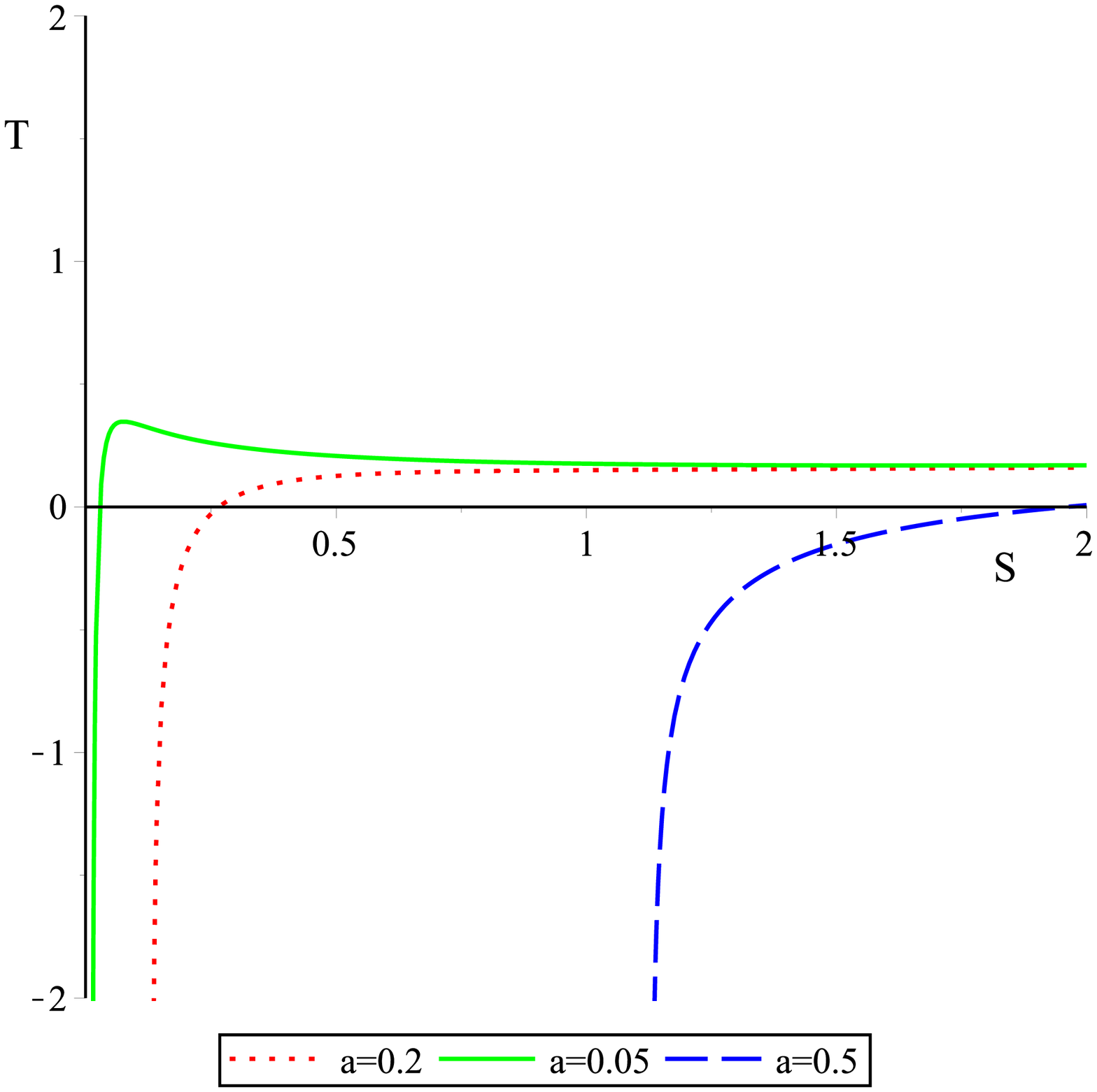,width=7cm}\caption{\small{Left plot: Temperature
$T$ with respect to entropy $S$ for $Q=0.05$, $a=0.5$, $\alpha=0.5$,
$\omega=-1$ and different values of $P$; Right plot: Temperature $T$
with respect to entropy $S$ for $Q=0.05$, $P=0.1$, $\alpha=0.5$,
$\omega=-1$ and different values of $a$ parameter.}}
\end{center}
\end{figure}
\begin{figure}
\hspace*{1cm}
\begin{center}
\epsfig{file=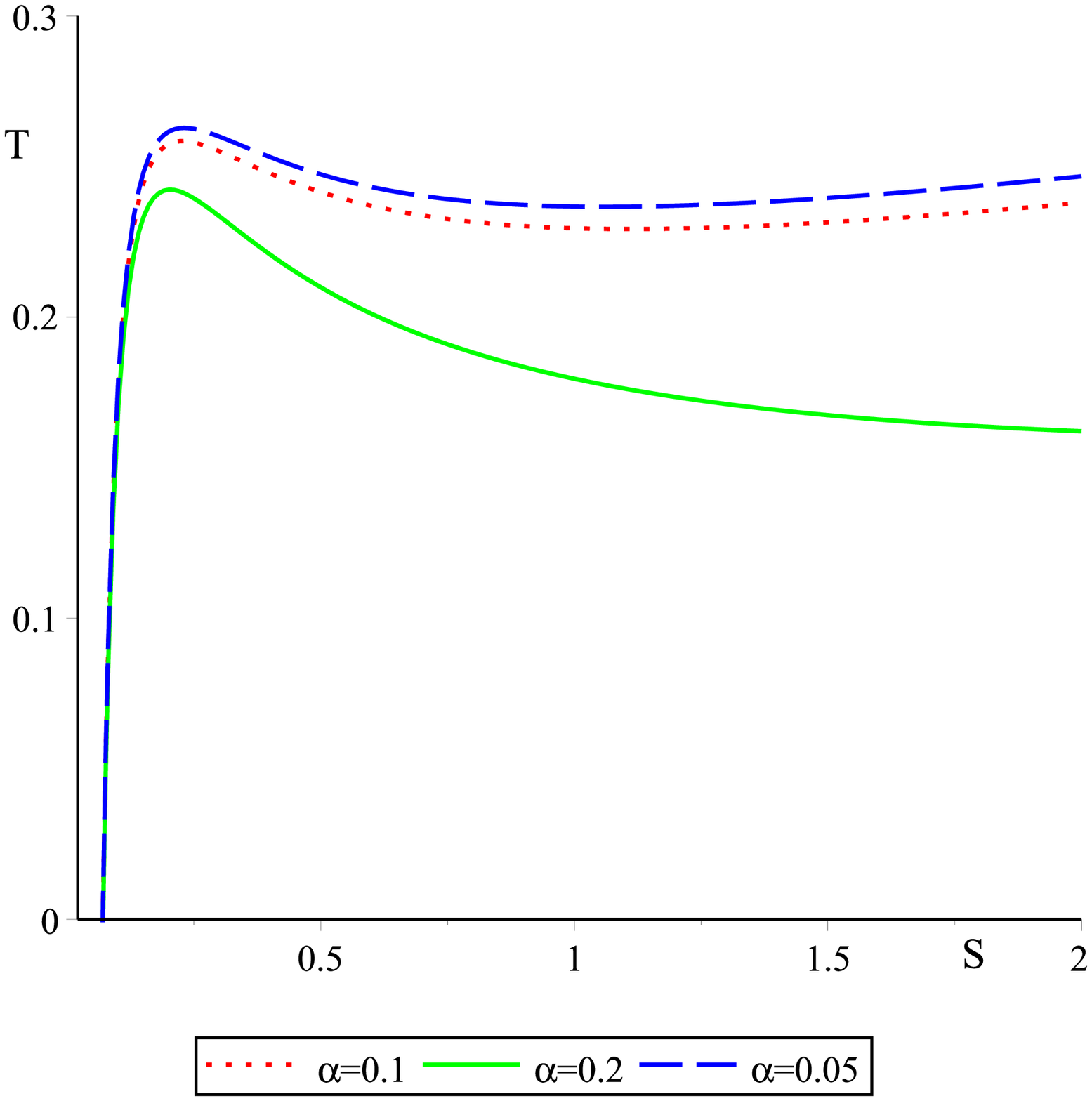,width=7cm}
\epsfig{file=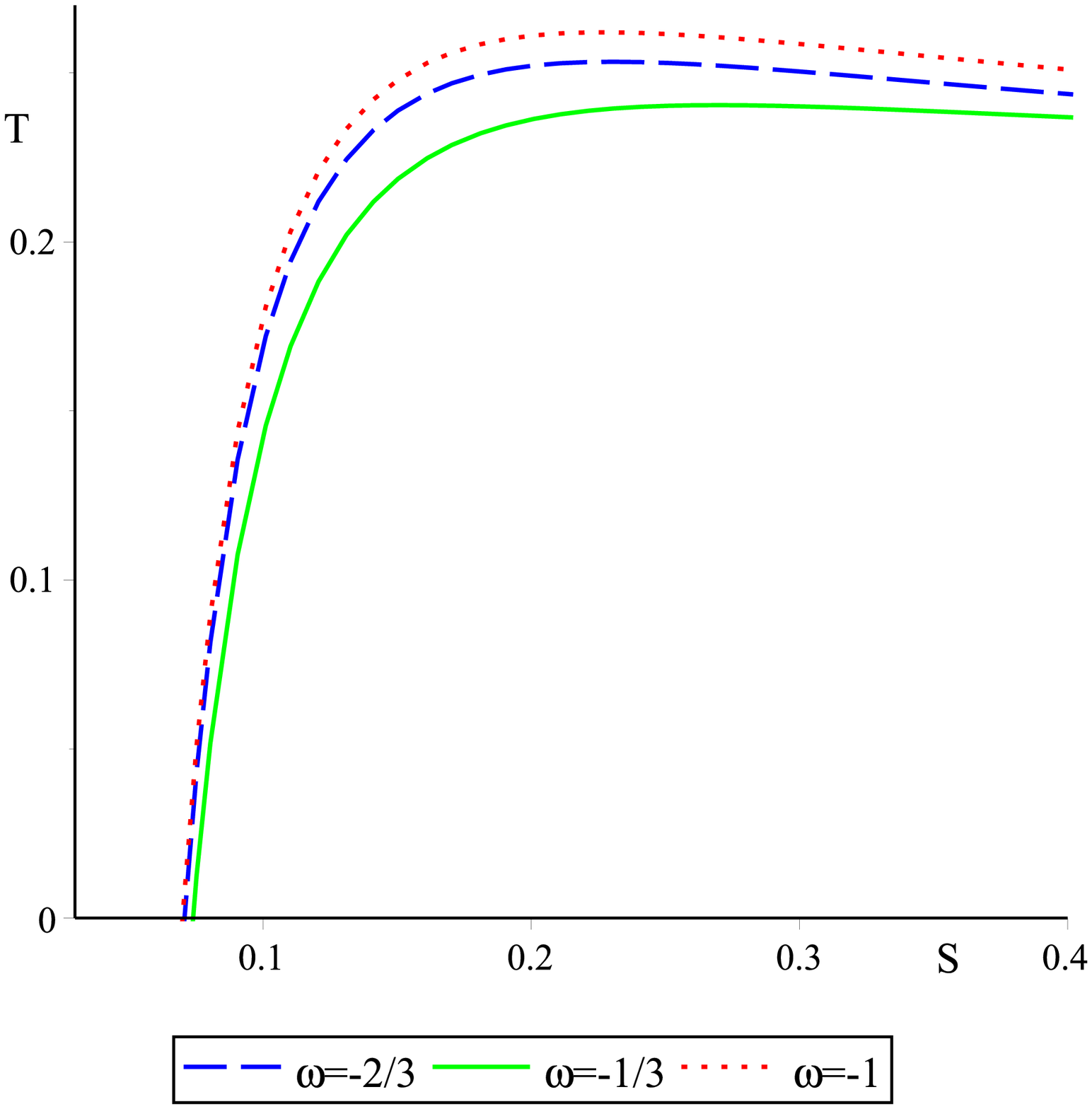,width=7cm}\caption{\small{Left plot: Temperature
$T$ with respect to entropy $S$ for $Q=0.05$, $a=0.1$, $P=0.1$,
$\omega=-1$ and different values of $\alpha$ parameter; Right plot:
Temperature $T$ with respect to entropy $S$ for $Q=0.05$, $P=0.1$,
$P=0.1$, $\alpha=0.1$ and different values of $\omega$ parameter.}}
\end{center}
\end{figure}

\section{Phase transition of Kerr-Newman-AdS black hole with quintessence}
In this section, we are going to study the phase transition of
Kerr-Newman-AdS black hole in Quintessential dark energy. The
parameter $\omega$ is an important parameter to determine the
property of spacetime metric. Here, we investigate two type of phase
transition for different values of $\omega$ and compare the obtained
result with each other. As mentioned in the introduction, the type
one of phase transition occurs when the temperature vanishes. We
plotted behavior of temperature with respect to entropy in figures
(1) and (2). As we see, type one of phase transition occurs for
$P<0.42$ and $a<0.5$ and the phase transition point shifts to higher
entropy when pressure $P$, rotation parameter $a$ and $\alpha$
increase. Also we see that by changing parameter $\omega$ from -1 to
$-\frac{1}{3}$, the critical point shifts to higher entropy.  We
need to calculate the heat capacity of the black hole to study type
two of phase transition. The heat capacity  is an interesting
thermodynamical quantity to determine stability and instability of
the black hole. As we know the heat capacity is negative for general
black hole then it is unstable and one can say a such black hole
produce Hawking radiation. But for charged and rotating black hole
the heat capacity could be positive and the black hole can have
phase transition. Now, we calculate heat capacity for different
$\omega$
and consider the phase transition.\\
\textbf{Case one:}$\boldsymbol{\omega=-\frac{1}{3}}$\\
The heat capacity is given by,
\begin{equation}
C=T\bigg(\frac{\partial S}{\partial T}\bigg)=T\bigg(\frac{\partial
S}{\partial r_{+}}\frac{\partial r_{+}}{\partial T}\bigg)
\end{equation}
Form equation (7), (9) and (11), we obtain heat capacity for
$\omega=-\frac{1}{3}$.

\begin{equation}
C=\frac{2\pi
r_{+}^{2}(r_{+}^{2}+a^{2})\bigg(\frac{3r_{+}^{2}}{\ell^{2}}+\frac{a^{2}}{\ell^{2}}+1-\frac{a^{2}+Q^{2}}{r_{+}^{2}}-\alpha
\bigg)}{\Xi\bigg(\frac{3r_{+}^{2}}{\ell^{2}}(r_{+}^{2}+3a^{2})+\frac{a^{2}+Q^{2}}{r_{+}^{2}}(3r_{+}^{2}+a^{2})+(a^{2}-r_{+}^{2})(\frac{a^{2}}{\ell^{2}}+1-\alpha)\bigg)}.
\end{equation}
One can rewrite equation (12) in terms of entropy $S$ and pressure
$P$ by substituting $r_{+}=\sqrt{\frac{S\Xi}{\pi}-a^{2}}$ and
$\ell^{2}=\frac{3}{8\pi P}$. In figures (3) and (4), we have plotted
heat capacity with respect to entropy. The left plot of Fig. (3)
show that heat capacity will be divergent for $P<0.35$ which is
representing type two of phase transition. As we see, there are two
critical points. For first case the system  translate to an unstable
phase afterward it will have a transition to a completely stability
phase. In right plot of fig. (3) we see that phase transition will
happen for $a<0.4$ and critical points increases when rotation
parameter decreases. By observing fig. (4) we notice that the system
will be divergent for $\alpha<0.2$ and $Q<0.1$. Also these two
figures show that the critical point shifts to higher entropy by
decreasing pressure, rotation and $\alpha$ parameters.

\begin{figure}
\hspace*{1cm}
\begin{center}
\epsfig{file=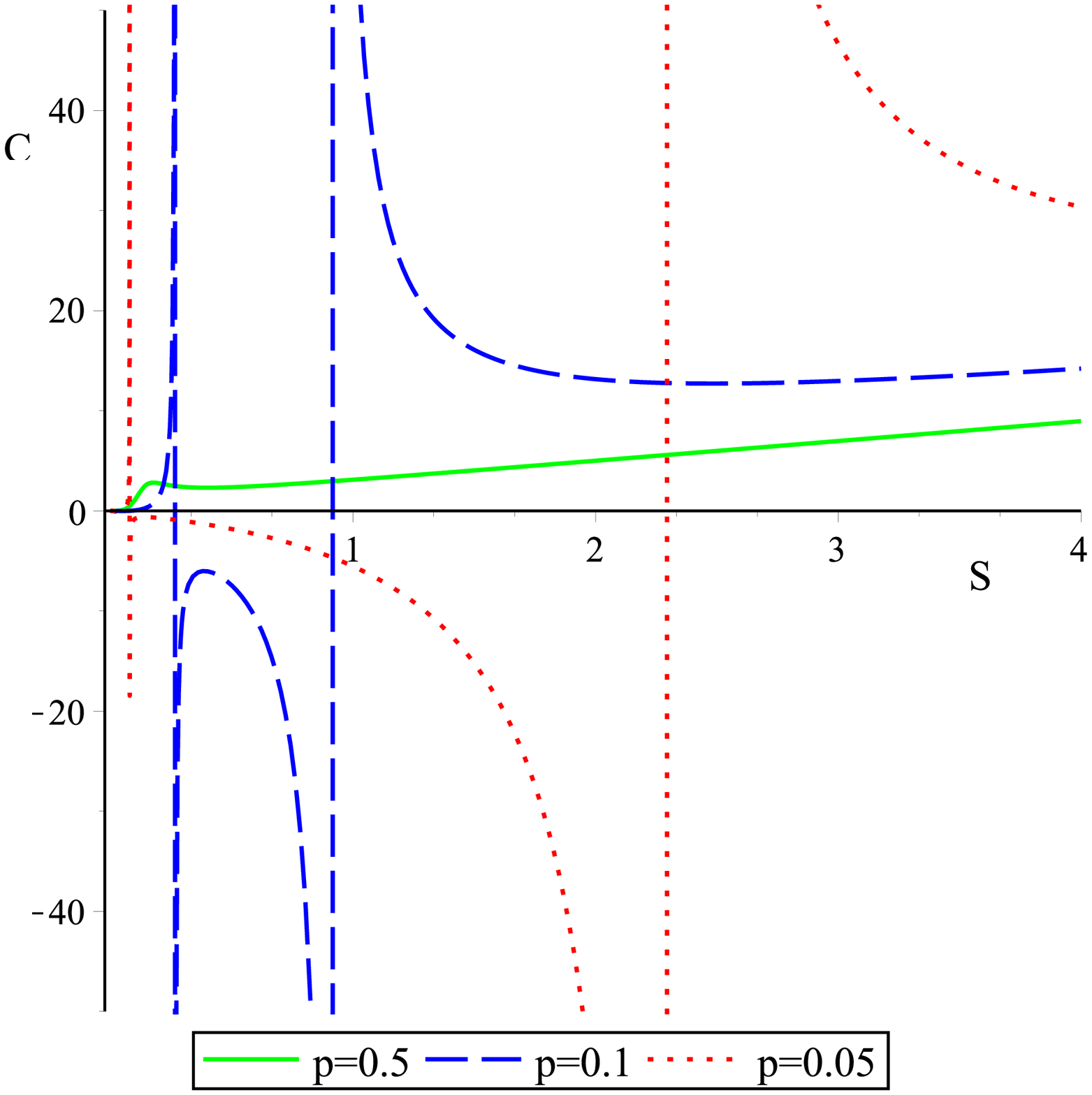,width=7cm}
\epsfig{file=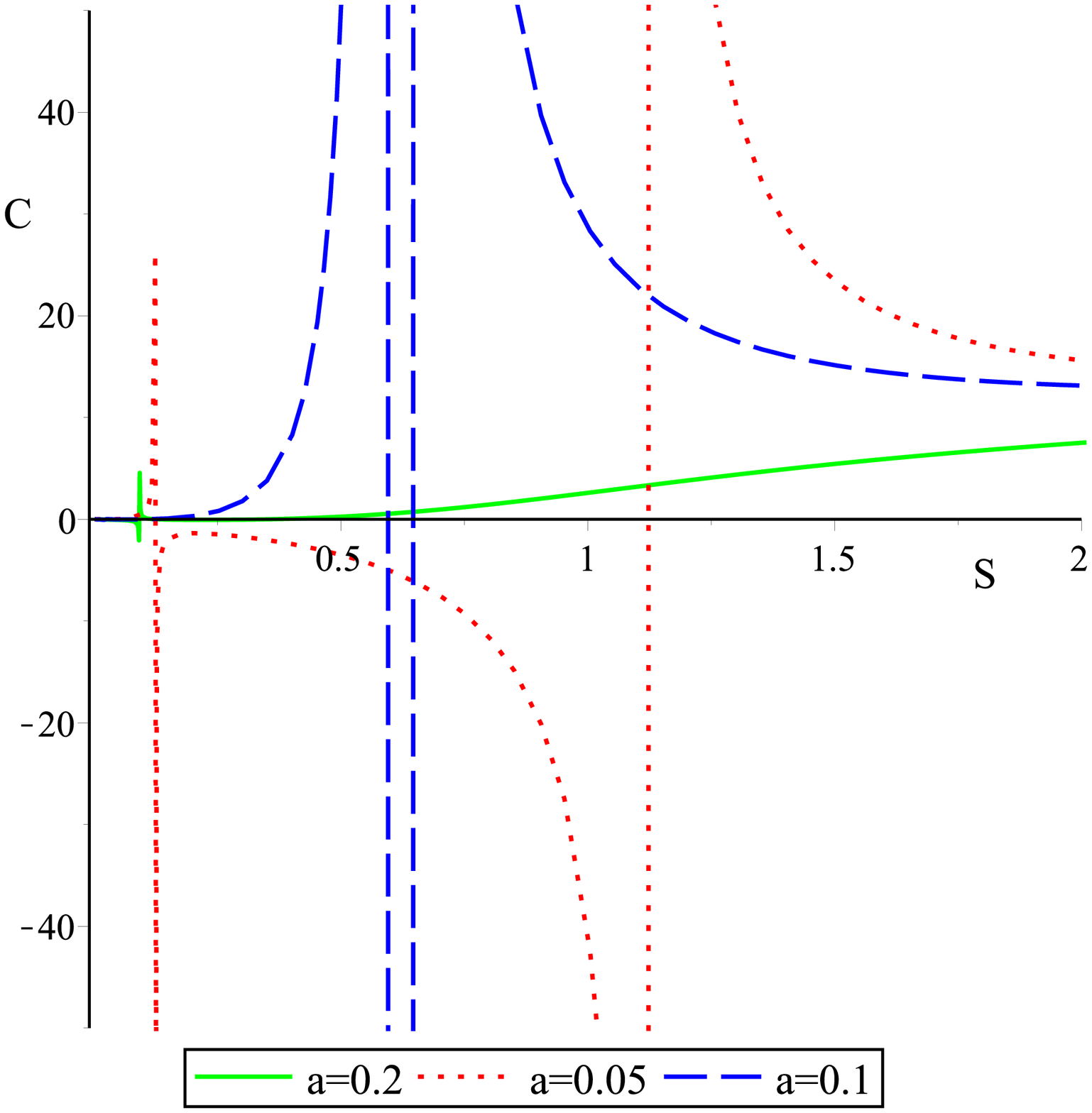,width=7cm}\caption{\small{Left plot: Heat
capacity $C$ with respect to entropy $S$ for $Q=0.05$, $a=0.1$,
$\alpha=0.5$, $\omega=-\frac{1}{3}$ and different values of $P$;
Right plot: Heat capacity $C$ with respect to entropy $S$ for
$Q=0.05$, $P=0.1$, $\alpha=0.05$, $\omega=-\frac{1}{3}$ and
different values of $a$.}}
\end{center}
\end{figure}

\begin{figure}
\hspace*{1cm}
\begin{center}
\epsfig{file=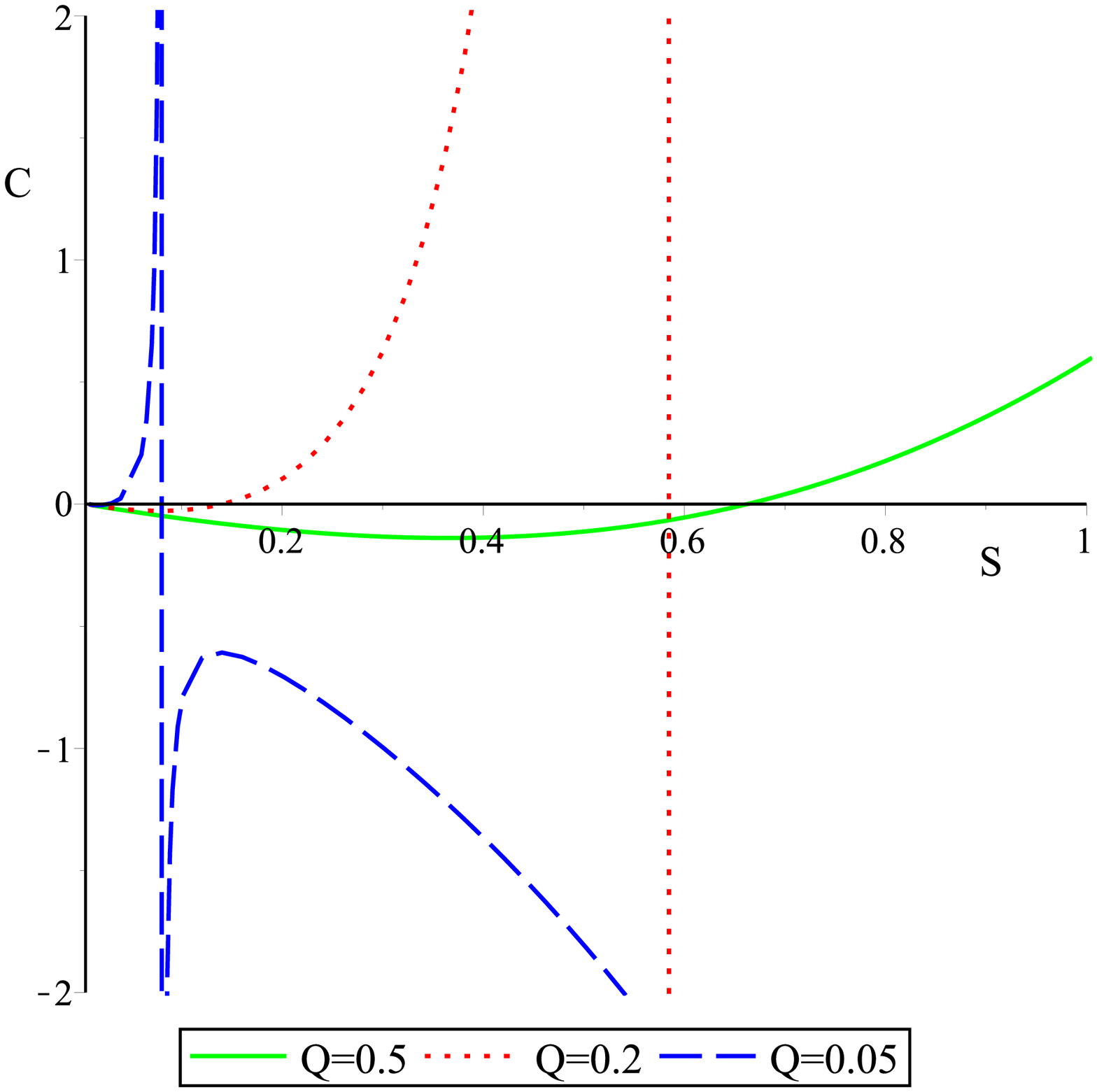,width=7cm}
\epsfig{file=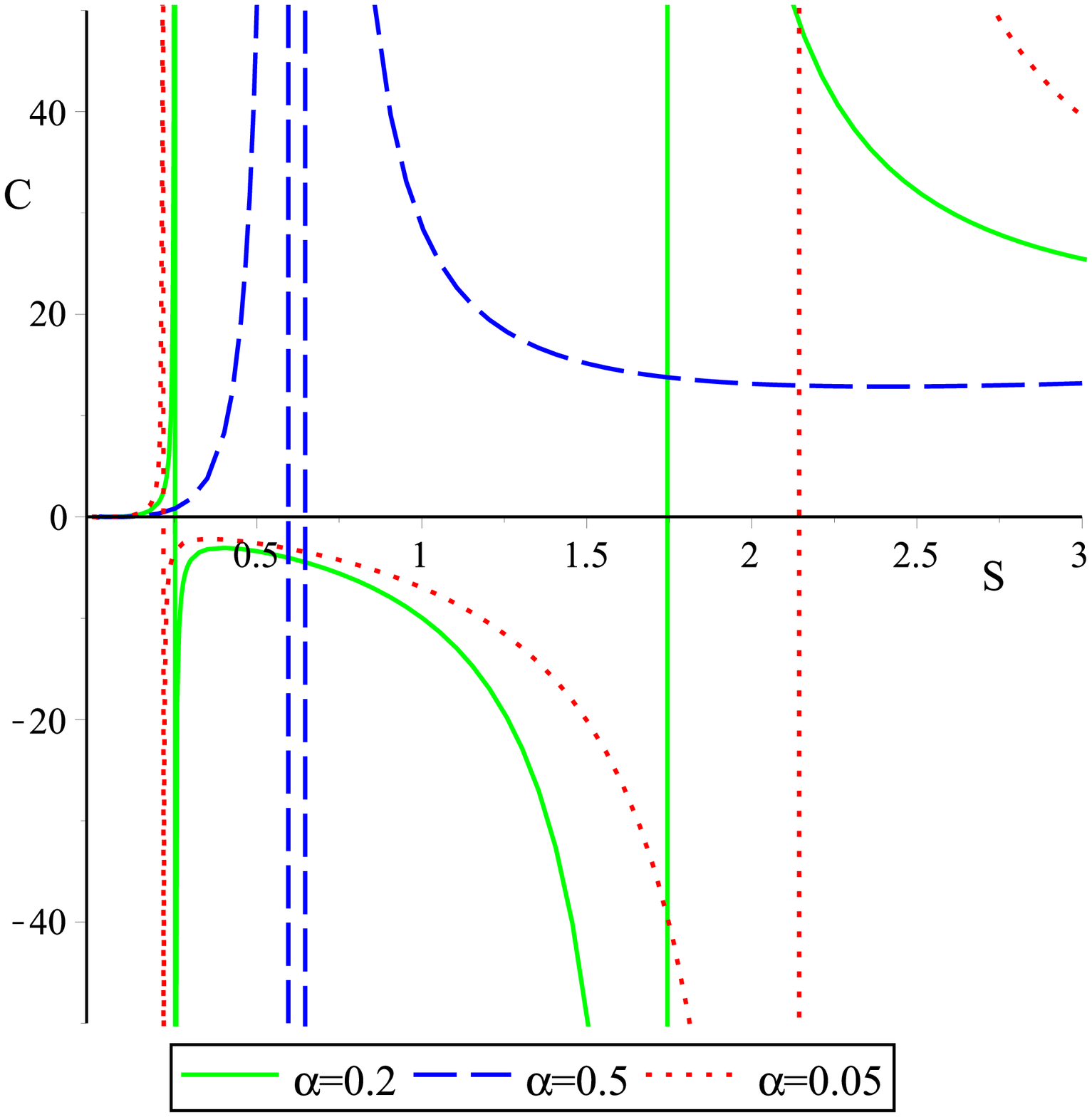,width=7cm}\caption{\small{Left plot: Heat
capacity $C$ with respect to entropy $S$ for $P=0.05$, $a=0.05$,
$\alpha=0.05$, $\omega=-\frac{1}{3}$ and different values of $Q$;
Right plot: Heat capacity $C$ with respect to entropy $S$ for
$Q=0.05$, $P=0.05$, $a=0.05$, $\omega=-\frac{1}{3}$ and different
values of $\alpha$.}}
\end{center}
\end{figure}

\textbf{Case two:} $\boldsymbol{\omega=-\frac{2}{3}}$\\
By substituting $\omega=-\frac{2}{3}$ in equation (9) and by
differentiating equation (7) and (9), we determine heat capacity as
follows,
\begin{equation}
C=\frac{2\pi
r_{+}^{2}(r_{+}^{2}+a^{2})\bigg(\frac{3r_{+}^{2}}{\ell^{2}}+\frac{a^{2}}{\ell^{2}}+1-\frac{a^{2}+Q^{2}}{r_{+}^{2}}-2\alpha
r_{+}
\bigg)}{\Xi\bigg(\frac{3r_{+}^{2}}{\ell^{2}}(r_{+}^{2}+3a^{2})+\frac{a^{2}+Q^{2}}{r_{+}^{2}}(3r_{+}^{2}+a^{2})+(a^{2}-r_{+}^{2})(\frac{a^{2}}{\ell^{2}}+1)-4\alpha
a^{2}r_{+}\bigg)}.
\end{equation}
In figures (5) and (6) we draw differentiation of heat capacity with
respect to entropy. Figure (5) show that heat capacity have two
divergent points for $P<0.15$ and $a<0.125$. As pervious case
($\omega=-\frac{1}{3}$), the first divergent point imply an unstable
transition and a stable transition occurs for second critical point.
Also we see that critical point shifts to lower entropy by
increasing pressure and rotation parameter. In left plot of fig. (6)
we see that the phase transition happen for $Q<0.1$ also right plot
show that $\alpha$ parameter does not affect on phase transition for
$\omega=-\frac{2}{3}$.

\begin{figure}
\hspace*{1cm}
\begin{center}
\epsfig{file=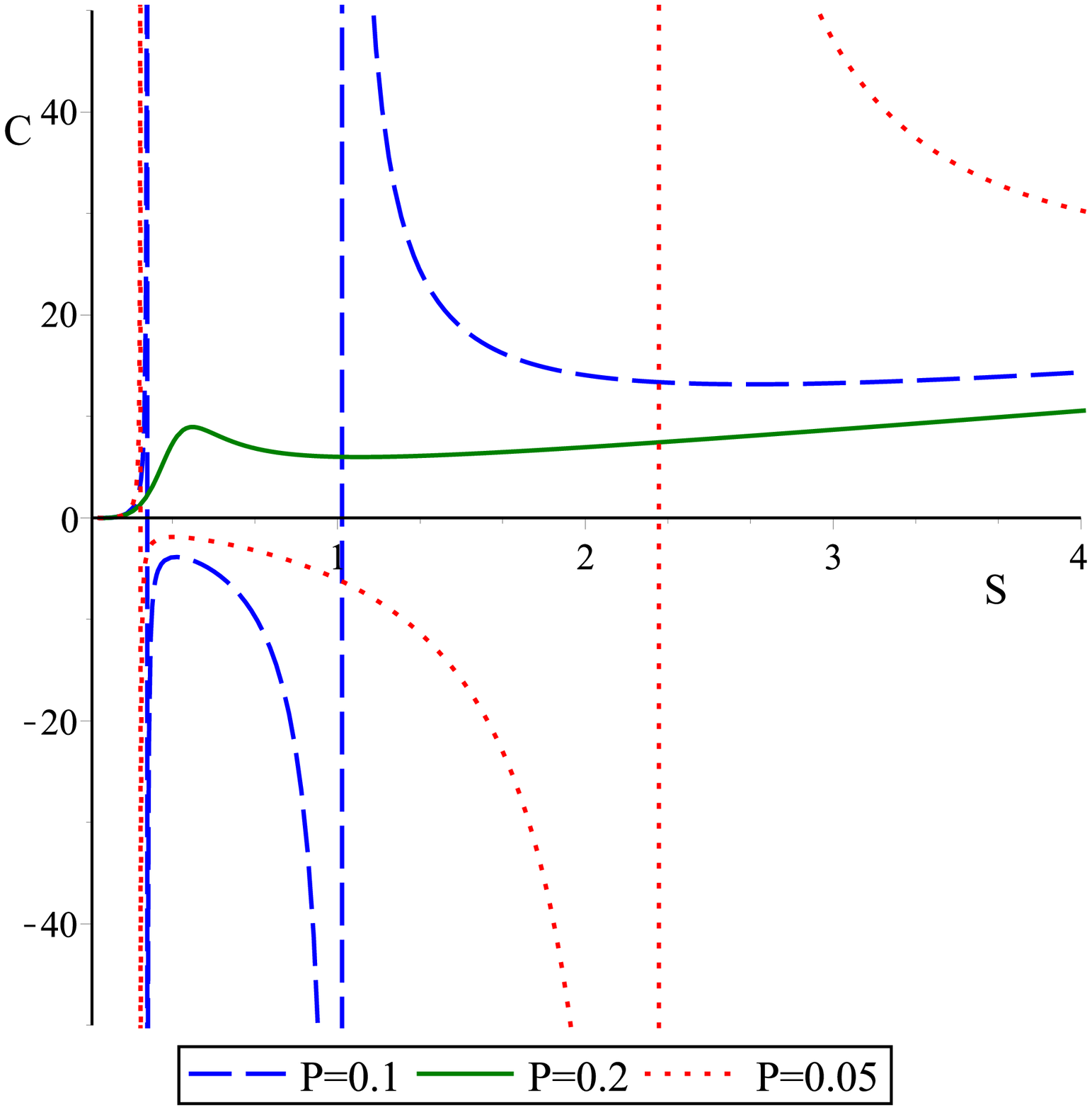,width=7cm}
\epsfig{file=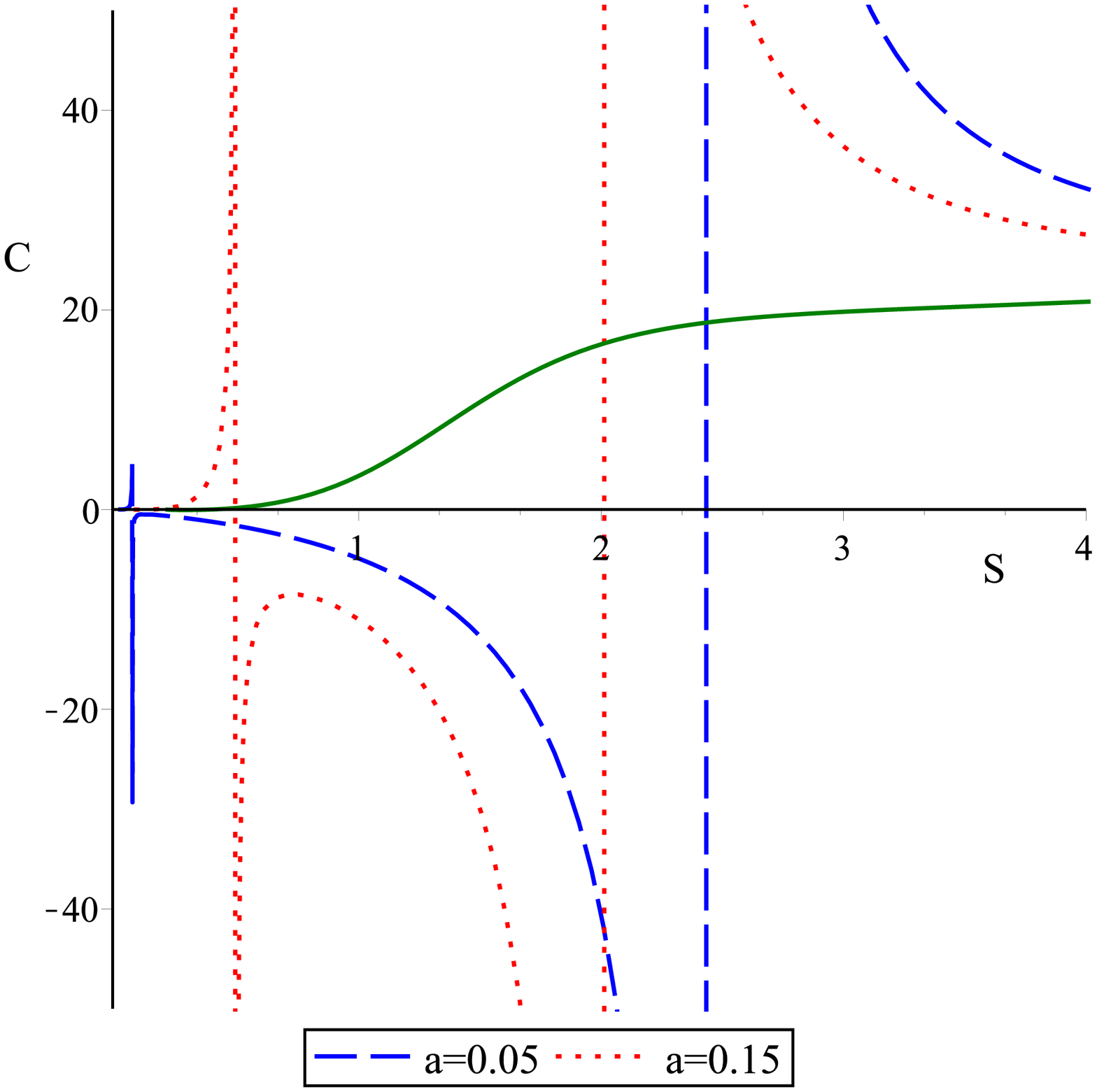,width=7cm}\caption{\small{Left plot: Heat
capacity $C$ with respect to entropy $S$ for $Q=0.05$, $a=0.1$,
$\alpha=0.5$, $\omega=-\frac{2}{3}$ and different values of $P$;
Right plot: Heat capacity $C$ with respect to entropy $S$ for
$Q=0.05$, $P=0.05$, $\alpha=0.05$, $\omega=-\frac{2}{3}$ and
different values of $a$.}}
\end{center}
\end{figure}

\begin{figure}
\hspace*{1cm}
\begin{center}
\epsfig{file=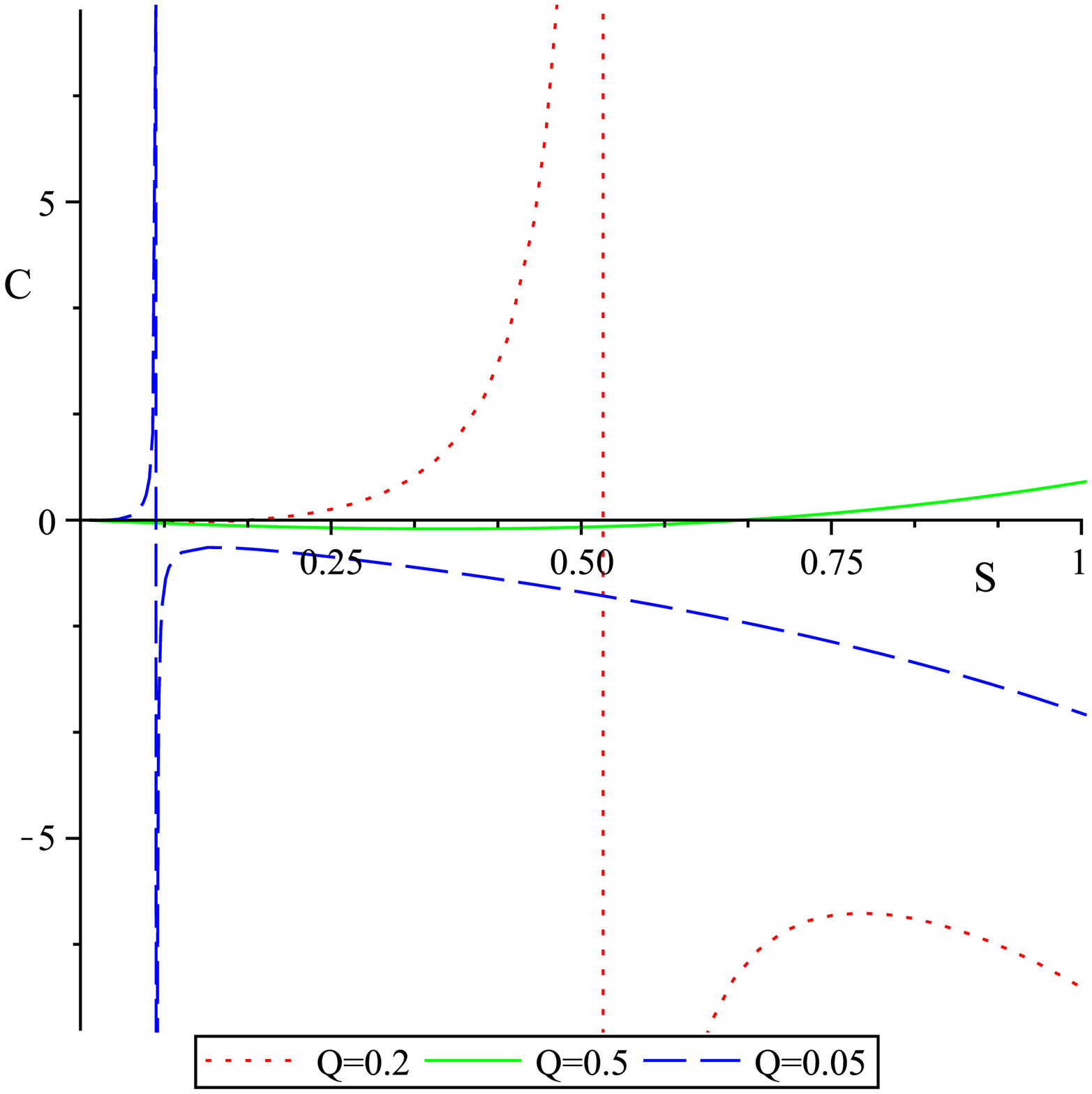,width=7cm}
\epsfig{file=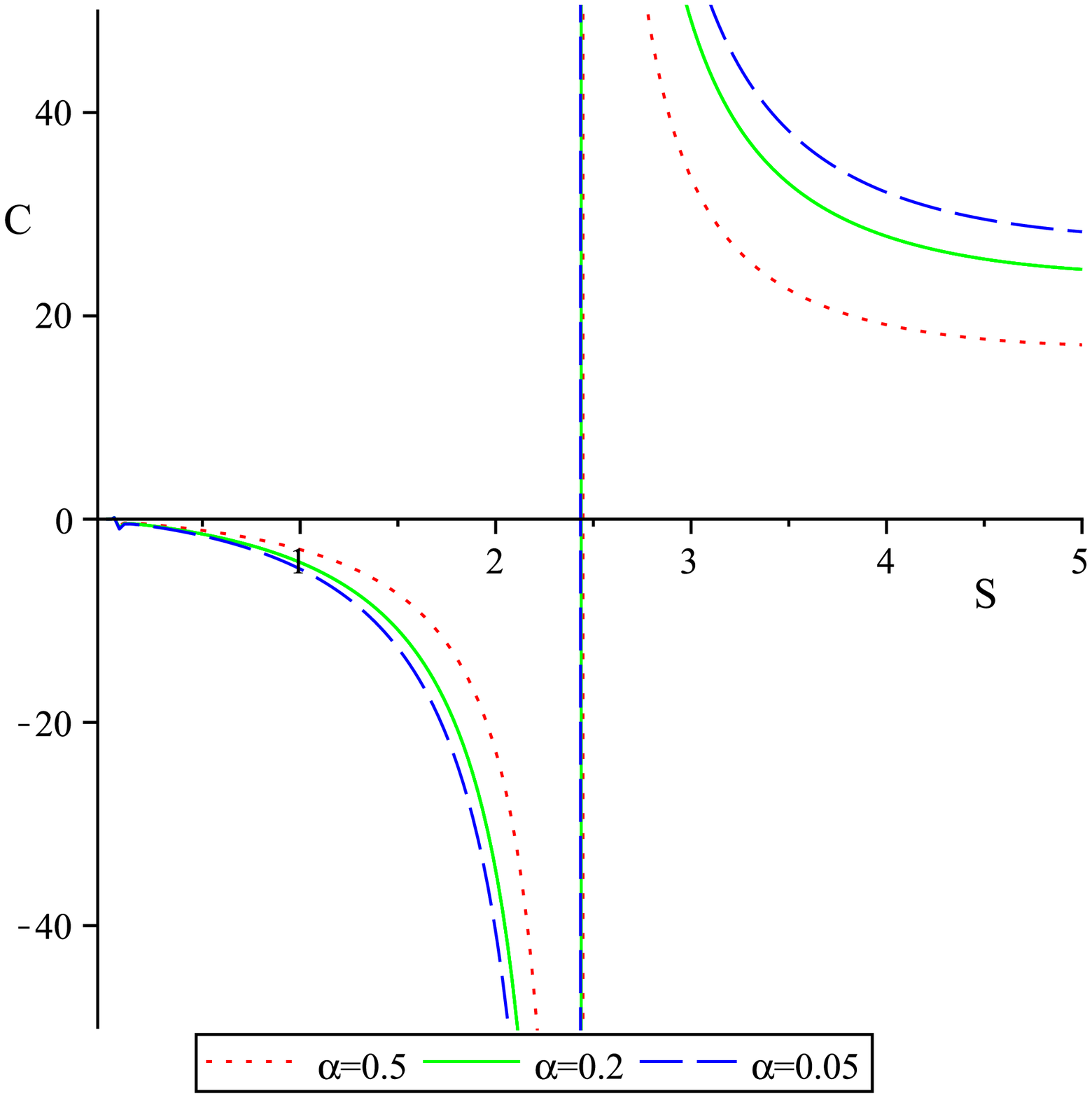,width=7cm}\caption{\small{Left plot: Heat
capacity $C$ with respect to entropy $S$ for $P=0.05$, $a=0.05$,
$\alpha=0.05$, $\omega=-\frac{2}{3}$ and different values of $Q$;
Right plot: Heat capacity $C$ with respect to entropy $S$ for
$Q=0.05$, $P=0.05$, $a=0.05$, $\omega=-\frac{2}{3}$ and different
values of $\alpha$.}}
\end{center}
\end{figure}

\textbf{Case three:}$\boldsymbol{\omega=-1}$\\
We can calculate heat capacity as previous steps. For this case, the
heat capacity is obtained as,
\begin{equation}
C=\frac{2\pi
r_{+}^{2}(r_{+}^{2}+a^{2})\bigg(\frac{3r_{+}^{2}}{\ell^{2}}+\frac{a^{2}}{\ell^{2}}+1-\frac{a^{2}+Q^{2}}{r_{+}^{2}}-3\alpha
r_{+}^{2} \bigg)}{\Xi\bigg((\frac{3r_{+}^{2}}{\ell^{2}}-3\alpha
r_{+}^{2})(r_{+}^{2}+3a^{2})+\frac{a^{2}+Q^{2}}{r_{+}^{2}}(3r_{+}^{2}+a^{2})+(a^{2}-r_{+}^{2})(\frac{a^{2}}{\ell^{2}}+1).
\bigg)}.
\end{equation}

We plot the behavior of the heat capacity with respect to entropy
$S$ in figures (7) and (8). By looking at Fig (7), we find that
phase transition will occurs for $P<0.45$ and $a\leq0.5$. And Fig
(4) show that the heat capacity will be divergent for $Q<0.1$ and
$\alpha\leq0.5$. Here, as pervious cases, the critical point move to
higher entropy by decreasing four parameters p, $a$, $\alpha$ and
$Q$. The only difference than two pervious cases is that the $C-Q$
plot has two divergent points.

\begin{figure}
\hspace*{1cm}
\begin{center}
\epsfig{file=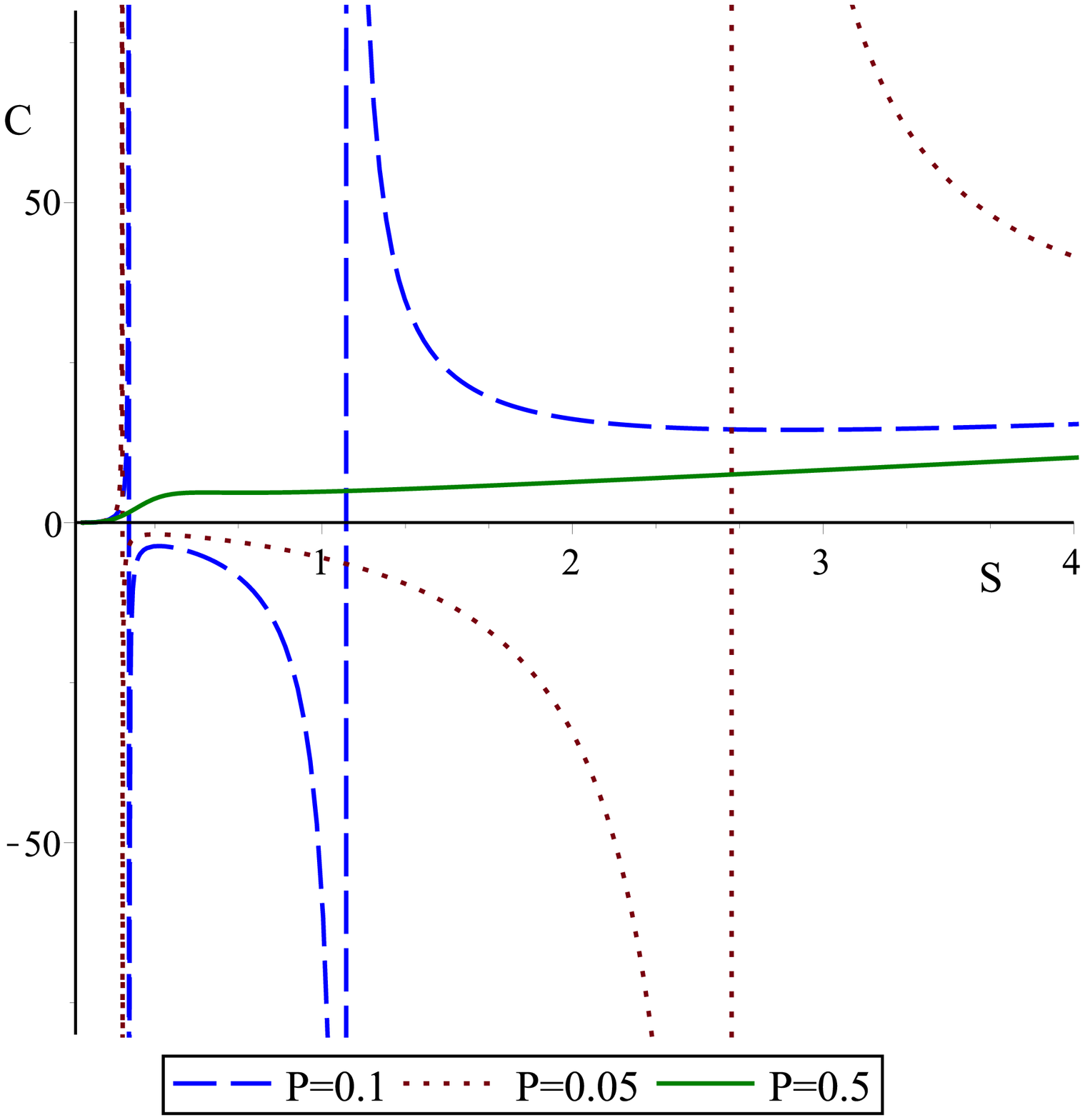,width=7cm}
\epsfig{file=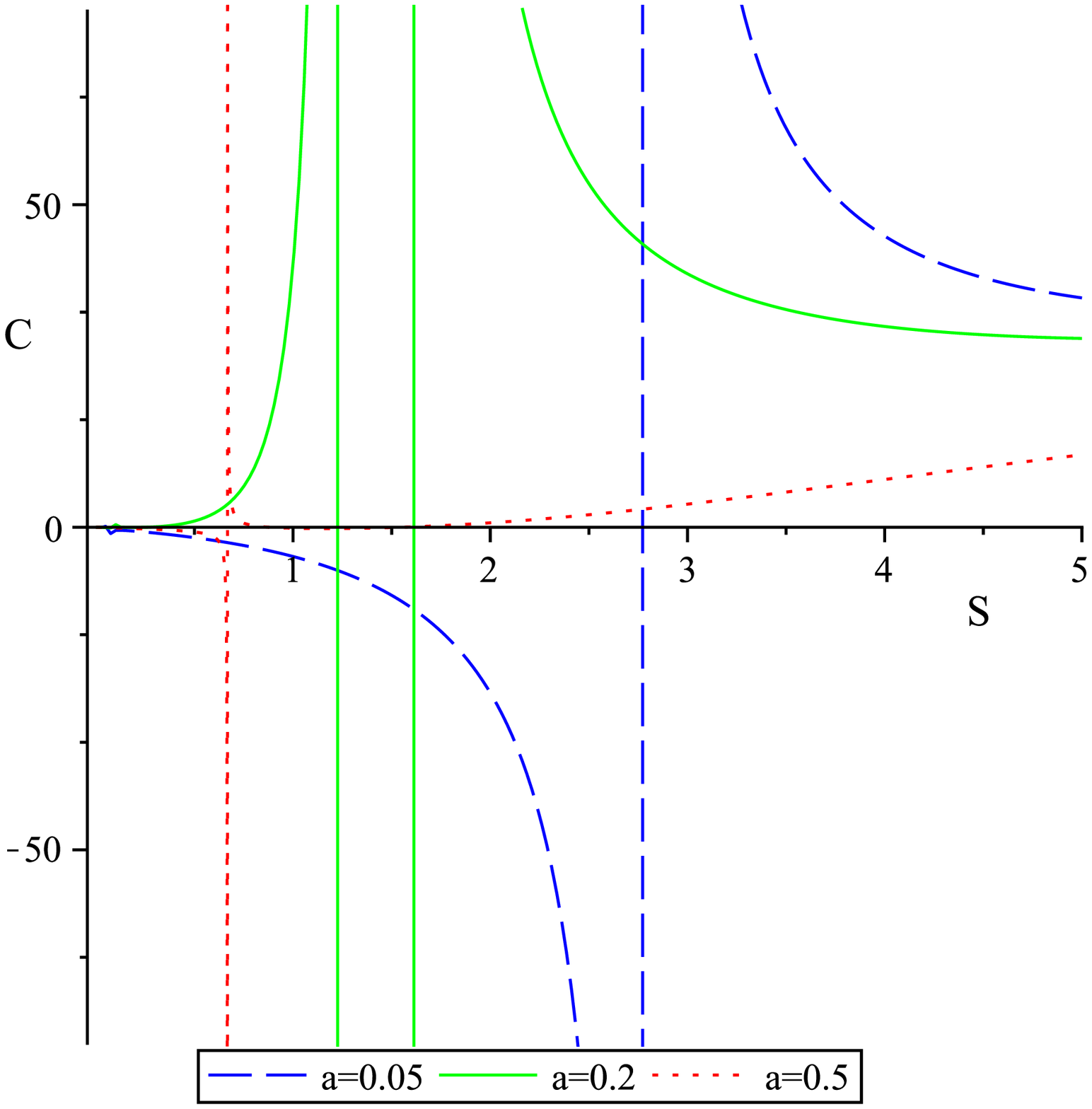,width=7cm}\caption{\small{Left plot: Heat
capacity $C$ with respect to entropy $S$ for $Q=0.05$, $a=0.1$,
$\alpha=0.5$, $\omega=-1$ and different values of $P$; Right plot:
Heat capacity $C$ with respect to entropy $S$ for $Q=0.05$,
$P=0.05$, $\alpha=0.05$, $\omega=-1$ and different values of $a$.}}
\end{center}
\end{figure}

\begin{figure}
\hspace*{1cm}
\begin{center}
\epsfig{file=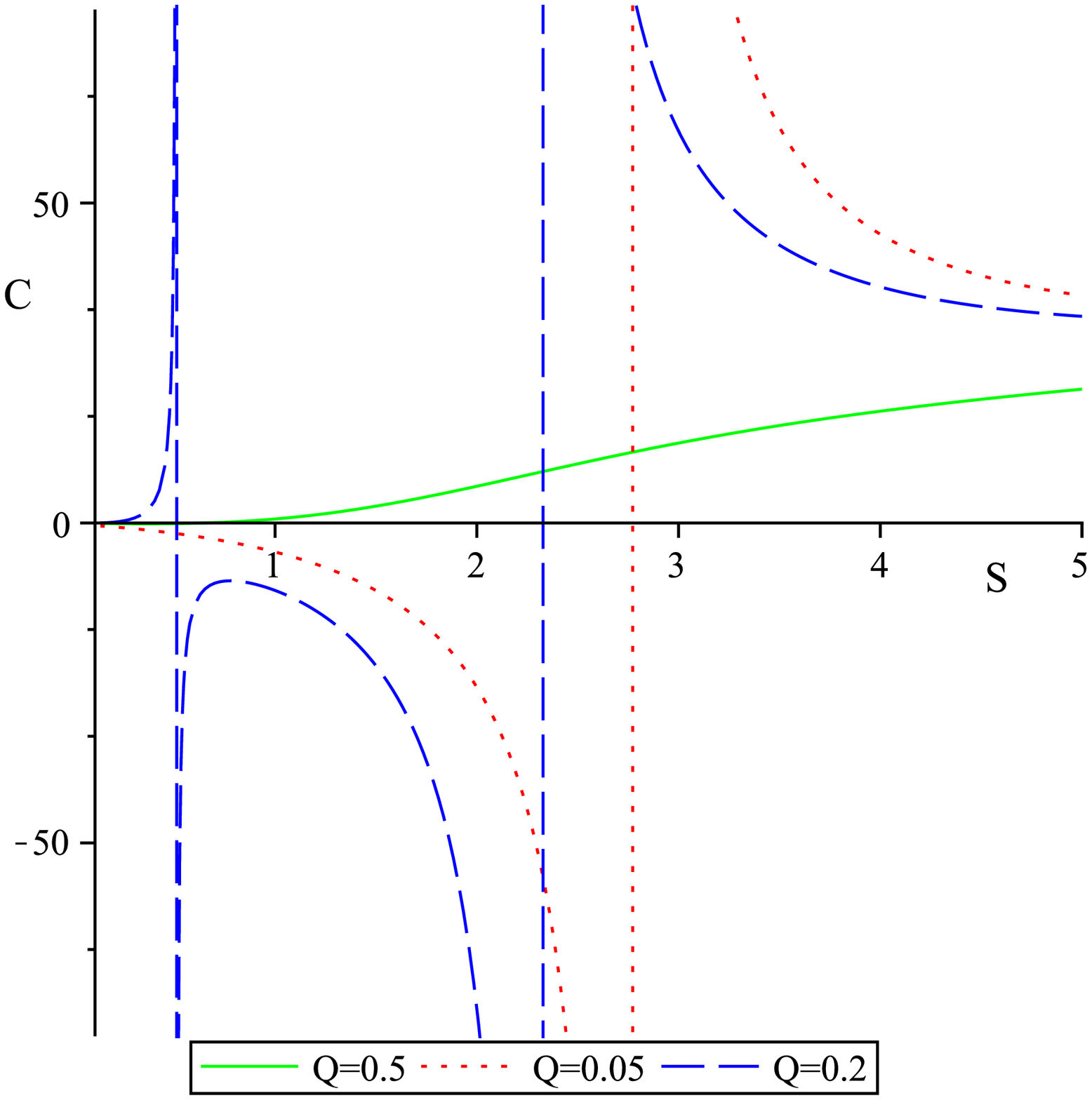,width=7cm}
\epsfig{file=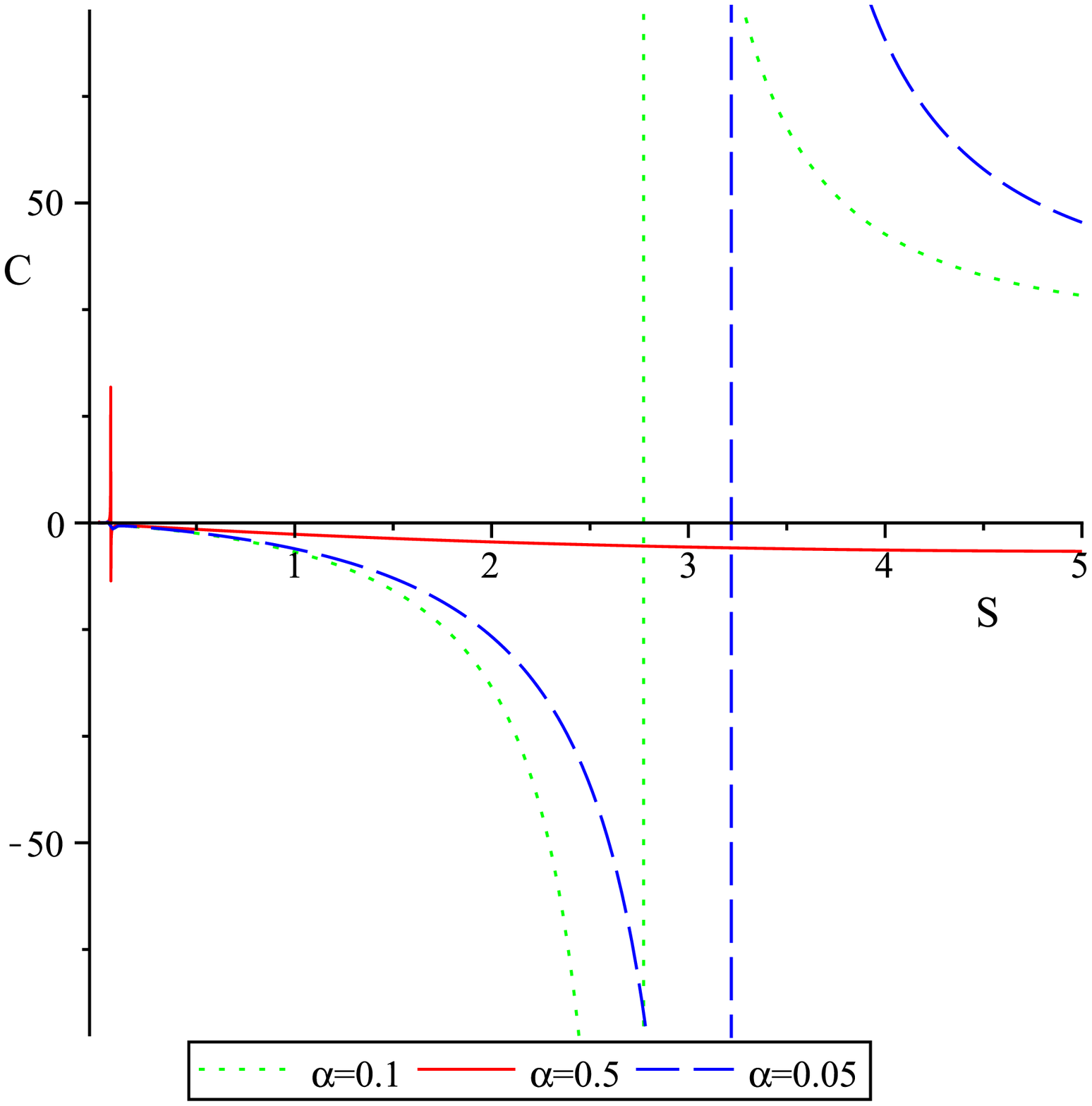,width=7cm}\caption{\small{Left plot: Heat
capacity $C$ with respect to entropy $S$ for $P=0.05$, $a=0.05$,
$\alpha=0.05$, $\omega=-1$ and different values of $Q$; Right plot:
Heat capacity $C$ with respect to entropy $S$ for $Q=0.05$,
$P=0.05$, $a=0.05$, $\omega=-1$ and different values of $\alpha$.}}
\end{center}
\end{figure}

\textbf{Case four:}$\boldsymbol{\omega=-\frac{1}{2}}$\\
The heat capacity is determined as follows,

\begin{equation}
C=\frac{2\pi
r_{+}^{2}(r_{+}^{2}+a^{2})\bigg(\frac{3r_{+}^{2}}{\ell^{2}}+\frac{a^{2}}{\ell^{2}}+1-\frac{a^{2}+Q^{2}}{r_{+}^{2}}-\frac{3}{2}\alpha
r_{+}^{\frac{1}{2}}
\bigg)}{\Xi\bigg(\frac{3r_{+}^{2}}{\ell^{2}}(r_{+}^{2}+3a^{2})+\frac{a^{2}+Q^{2}}{r_{+}^{2}}(3r_{+}^{2}+a^{2})+(a^{2}-r_{+}^{2})(\frac{a^{2}}{\ell^{2}}+1)-\frac{3}{4}\alpha
r_{+}^{\frac{1}{2}}(3a^{2}-r_{+}^{2})\bigg)}.
\end{equation}
We plot the heat capacity behavior with respect to entropy in
figures (9) and (10). Fig. (9) show that the system will have a
phase transition for $P<0.08$ and $a<0.125$. In left plot of fig.
(10), we see that there are two critical points for $Q\leq0.2$. This
plot is similar to left plot of fig. (8)  for $\omega=-1$ but here
critical points occur in lower entropy. As we see in right plot of
fig. (10), the heat capacity will be diverge for $\alpha\leq0.5$ and
also we see that critical points decrease by decreasing $\alpha$
parameter.

\begin{figure}
\hspace*{1cm}
\begin{center}
\epsfig{file=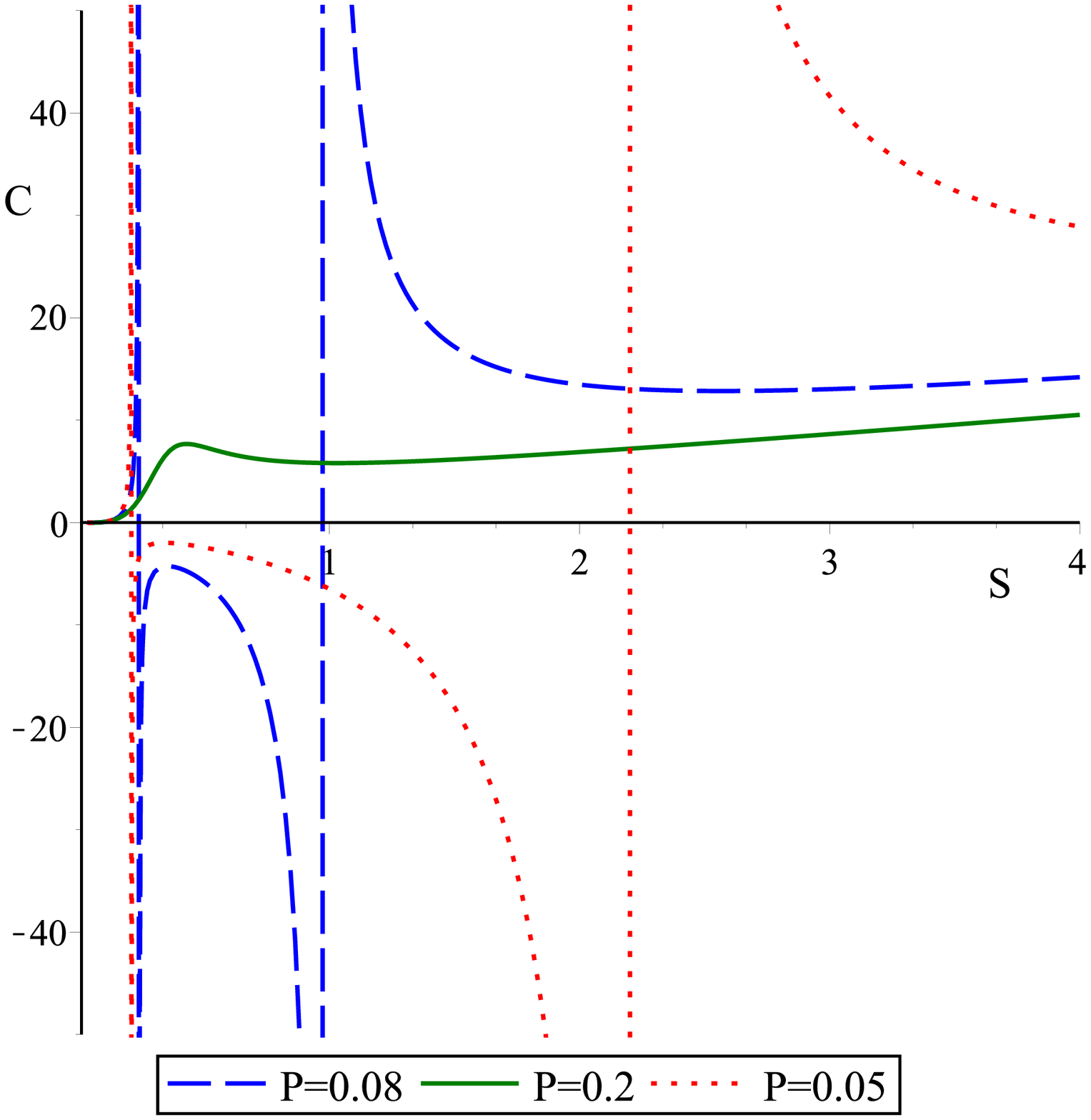,width=7cm}
\epsfig{file=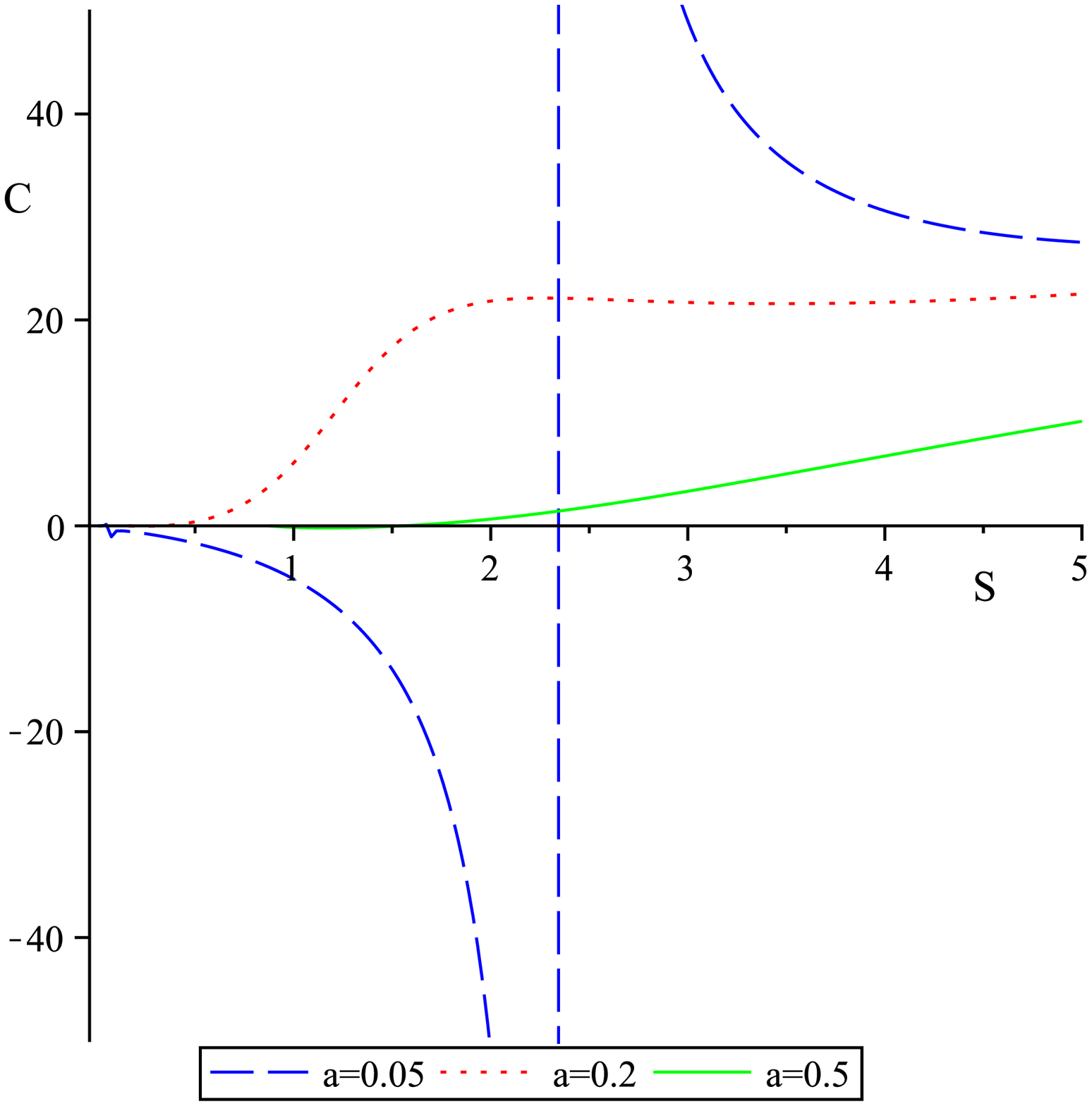,width=7cm}\caption{\small{Left plot: Heat
capacity $C$ with respect to entropy $S$ for $Q=0.05$, $a=0.1$,
$\alpha=0.5$, $\omega=-\frac{1}{2}$ and different values of $P$;
Right plot: Heat capacity $C$ with respect to entropy $S$ for
$Q=0.05$, $P=0.05$, $\alpha=0.05$, $\omega=-\frac{1}{2}$ and
different values of $a$.}}
\end{center}
\end{figure}

\begin{figure}
\hspace*{1cm}
\begin{center}
\epsfig{file=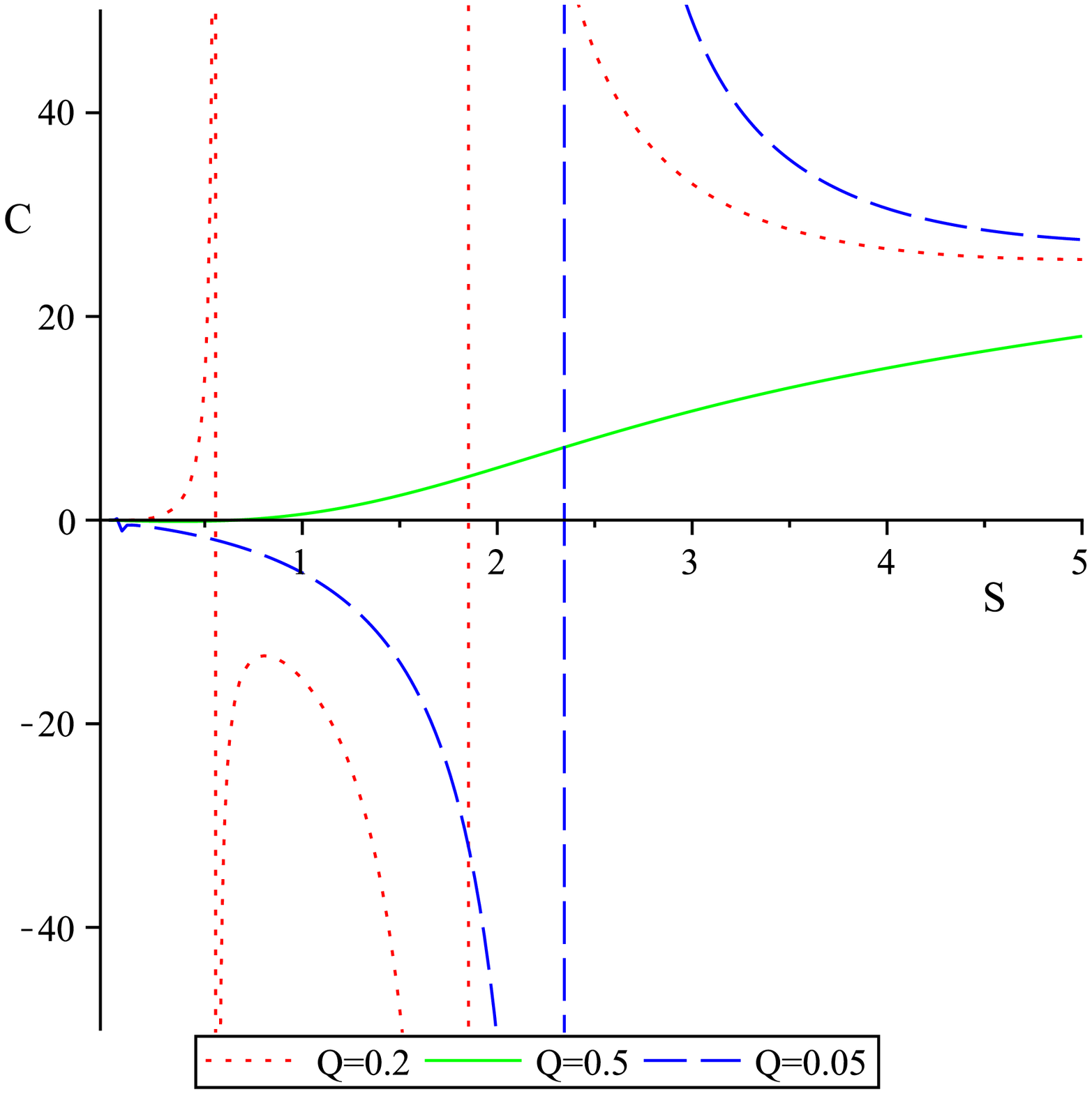,width=7cm}
\epsfig{file=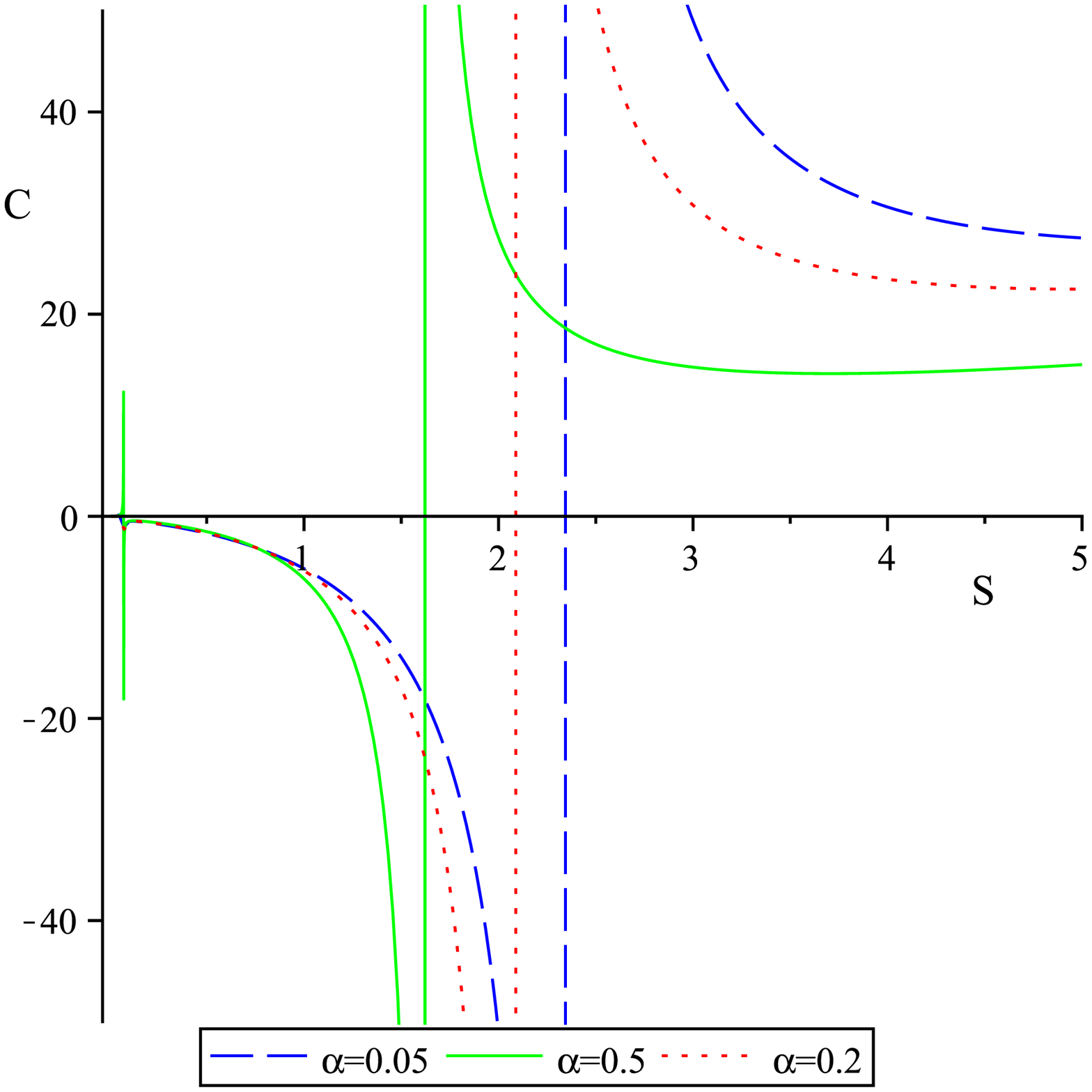,width=7cm}\caption{\small{Left plot: Heat
capacity $C$ with respect to entropy $S$ for $P=0.05$, $a=0.05$,
$\alpha=0.05$, $\omega=-\frac{1}{2}$ and different values of $Q$;
Right plot: Heat capacity $C$ with respect to entropy $S$ for
$Q=0.05$, $P=0.05$, $a=0.05$, $\omega=-\frac{1}{2}$ and different
values of $\alpha$.}}
\end{center}
\end{figure}

\textbf{Case five:} $\boldsymbol{\omega=-\frac{2}{5}}$\\
The heat capacity is given by,
\begin{equation}
C=\frac{2\pi
r_{+}^{2}(r_{+}^{2}+a^{2})\bigg(\frac{3r_{+}^{2}}{\ell^{2}}+\frac{a^{2}}{\ell^{2}}+1-\frac{a^{2}+Q^{2}}{r_{+}^{2}}-\frac{6}{5}\alpha
r_{+}^{\frac{1}{5}}
\bigg)}{\Xi\bigg(\frac{3r_{+}^{2}}{\ell^{2}}(r_{+}^{2}+3a^{2})+\frac{a^{2}+Q^{2}}{r_{+}^{2}}(3r_{+}^{2}+a^{2})+(a^{2}-r_{+}^{2})(\frac{a^{2}}{\ell^{2}}+1)-\frac{12}{25}\alpha
r_{+}^{\frac{1}{5}}(3a^{2}-2r_{+}^{2})\bigg)}.
\end{equation}
We draw $C - S$ diagram in figures (11) and (12). From Fig.(11), we
notice that phase transition will happen for $P<0.055$ and $a<0.1$
and also right plot of figure is similar to right plot of fig. (9).
And left plot of Fig.(11) show that phase transition exist for
$Q<0.2$ and also this plot is alike to left plot of fig. (4). The
difference is that critical point occurs in lower entropy here. By
observing the right plot of figure (12), we find that the heat
capacity will be diverge for $\alpha\leq0.5$. And also divergent
points decrease by decreasing $\alpha$ parameter which is similar to
right plot of figure (10).

\begin{figure}
\hspace*{1cm}
\begin{center}
\epsfig{file=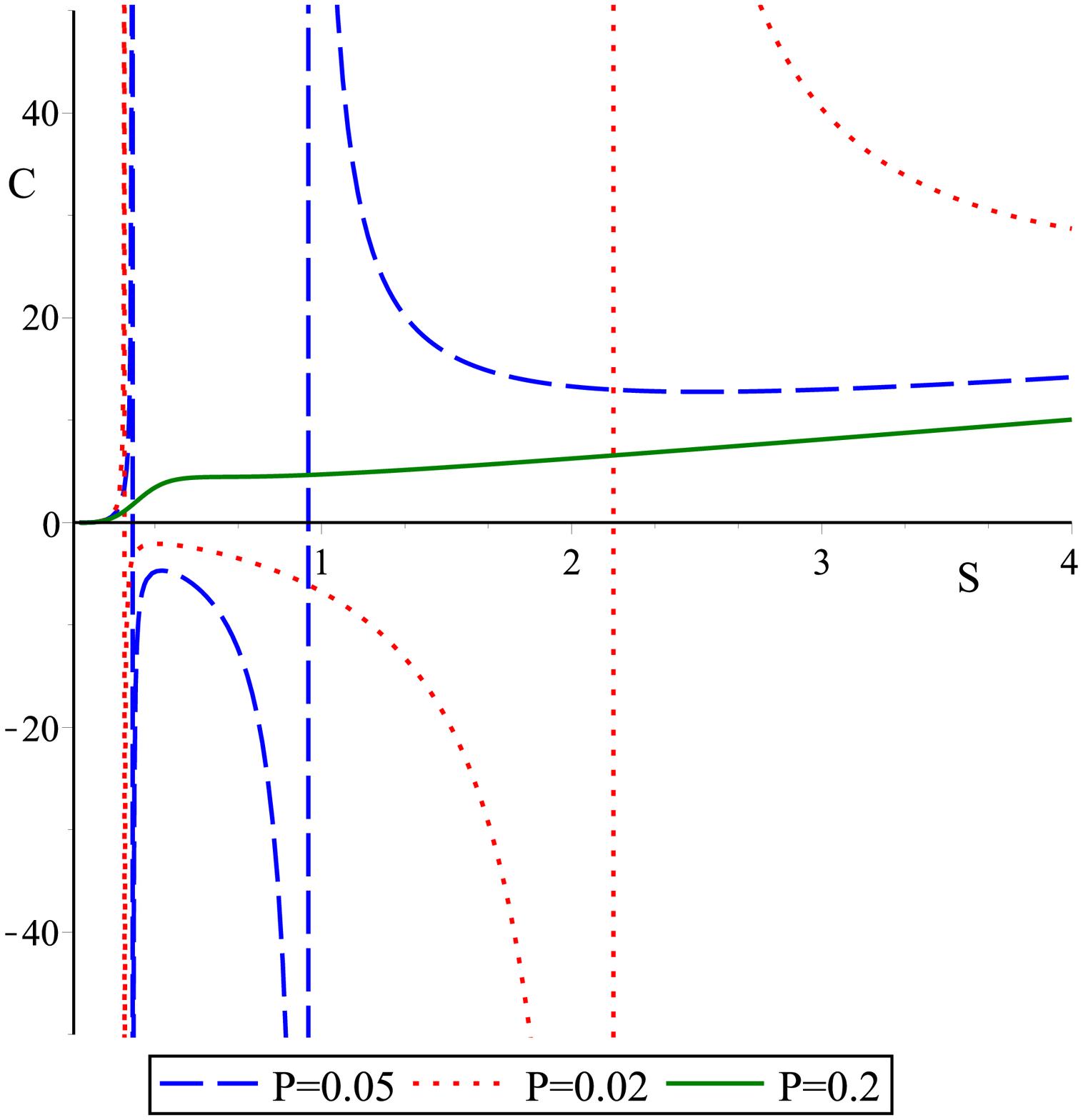,width=7cm}
\epsfig{file=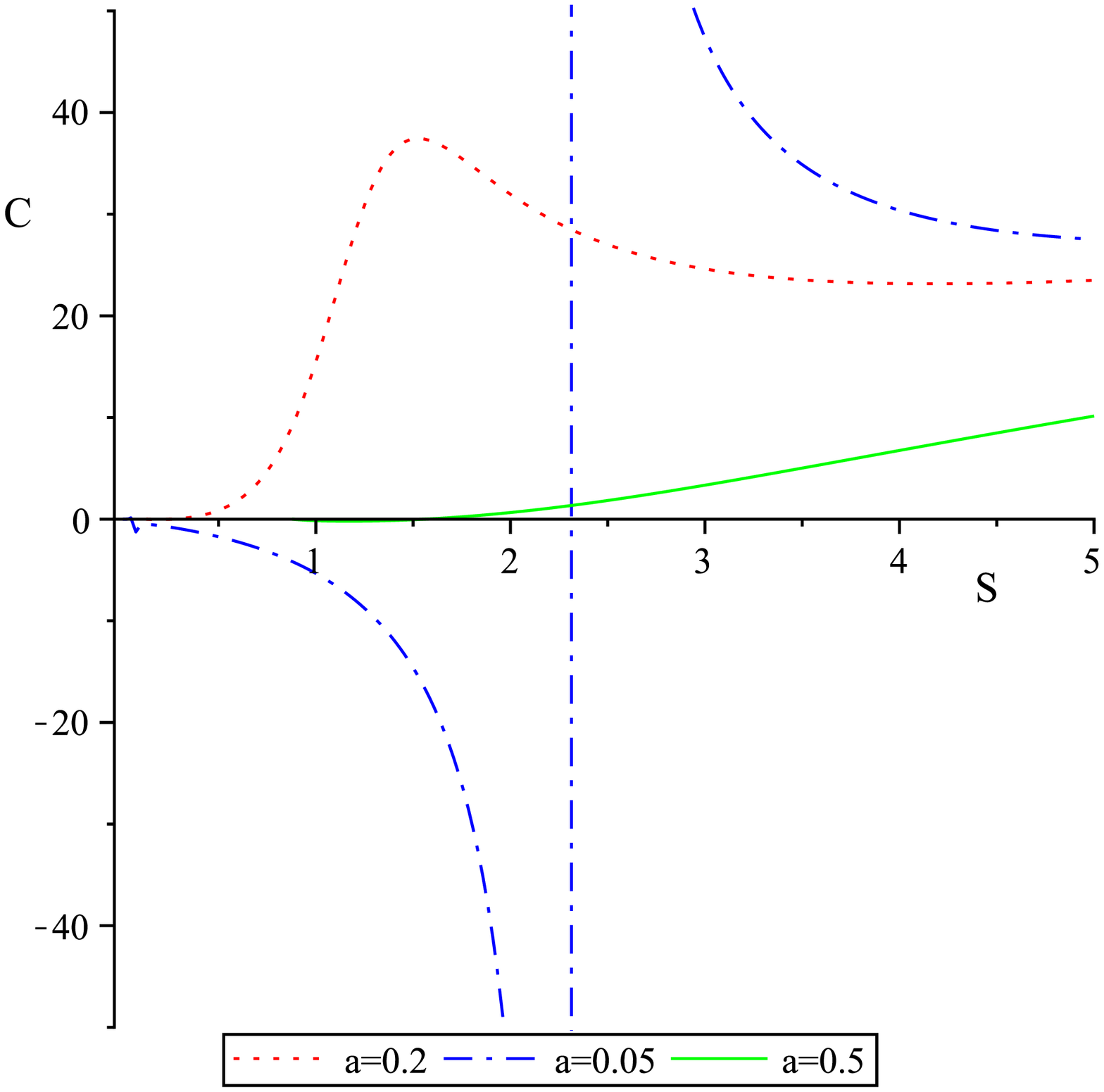,width=7cm}\caption{\small{Left plot: Heat
capacity $C$ with respect to entropy $S$ for $Q=0.05$, $a=0.1$,
$\alpha=0.5$, $\omega=-\frac{2}{5}$ and different values of $P$;
Right plot: Heat capacity $C$ with respect to entropy $S$ for
$Q=0.05$, $P=0.05$, $\alpha=0.05$, $\omega=-\frac{2}{5}$ and
different values of $a$.}}
\end{center}
\end{figure}

\begin{figure}
\hspace*{1cm}
\begin{center}
\epsfig{file=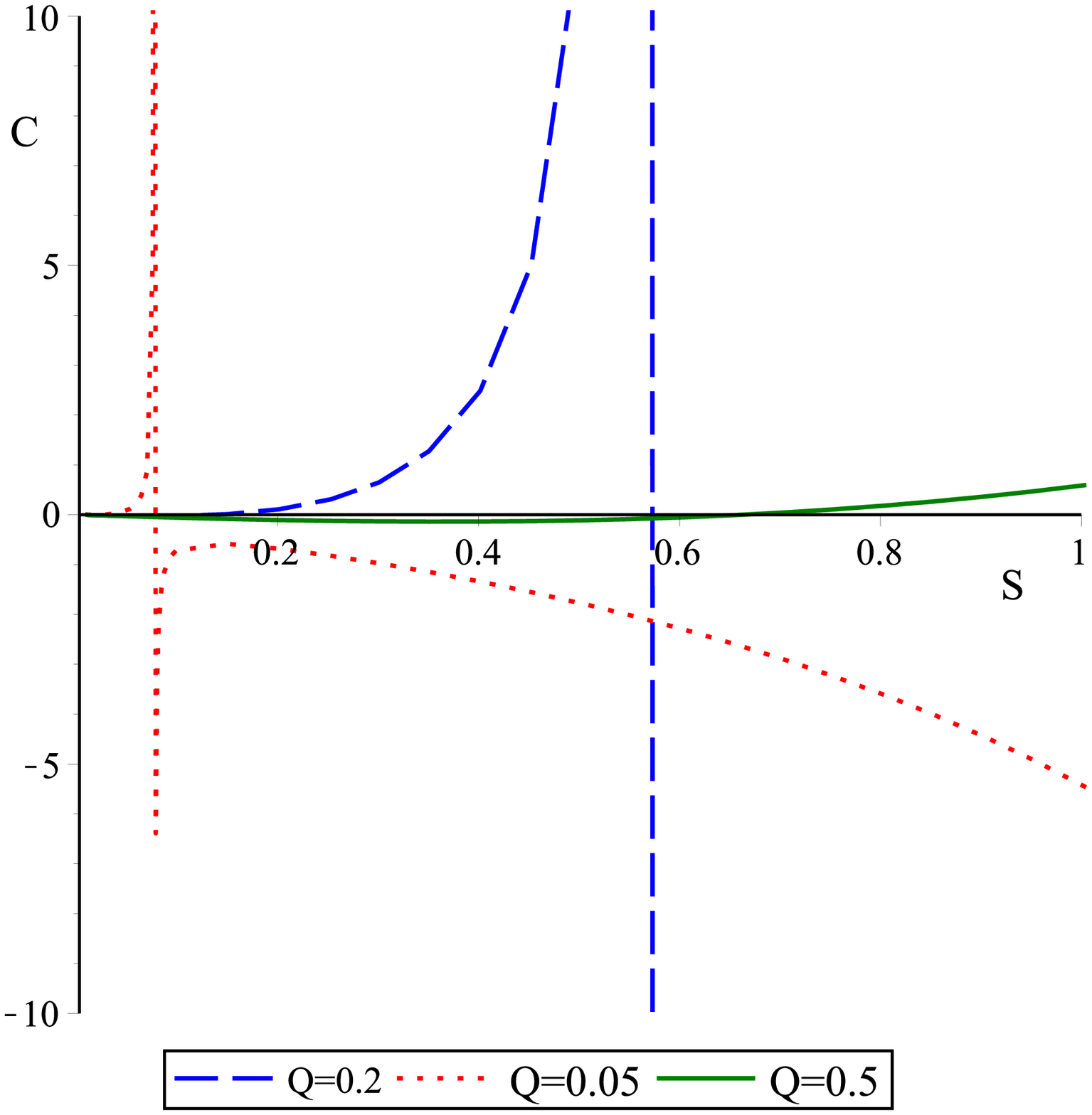,width=7cm}
\epsfig{file=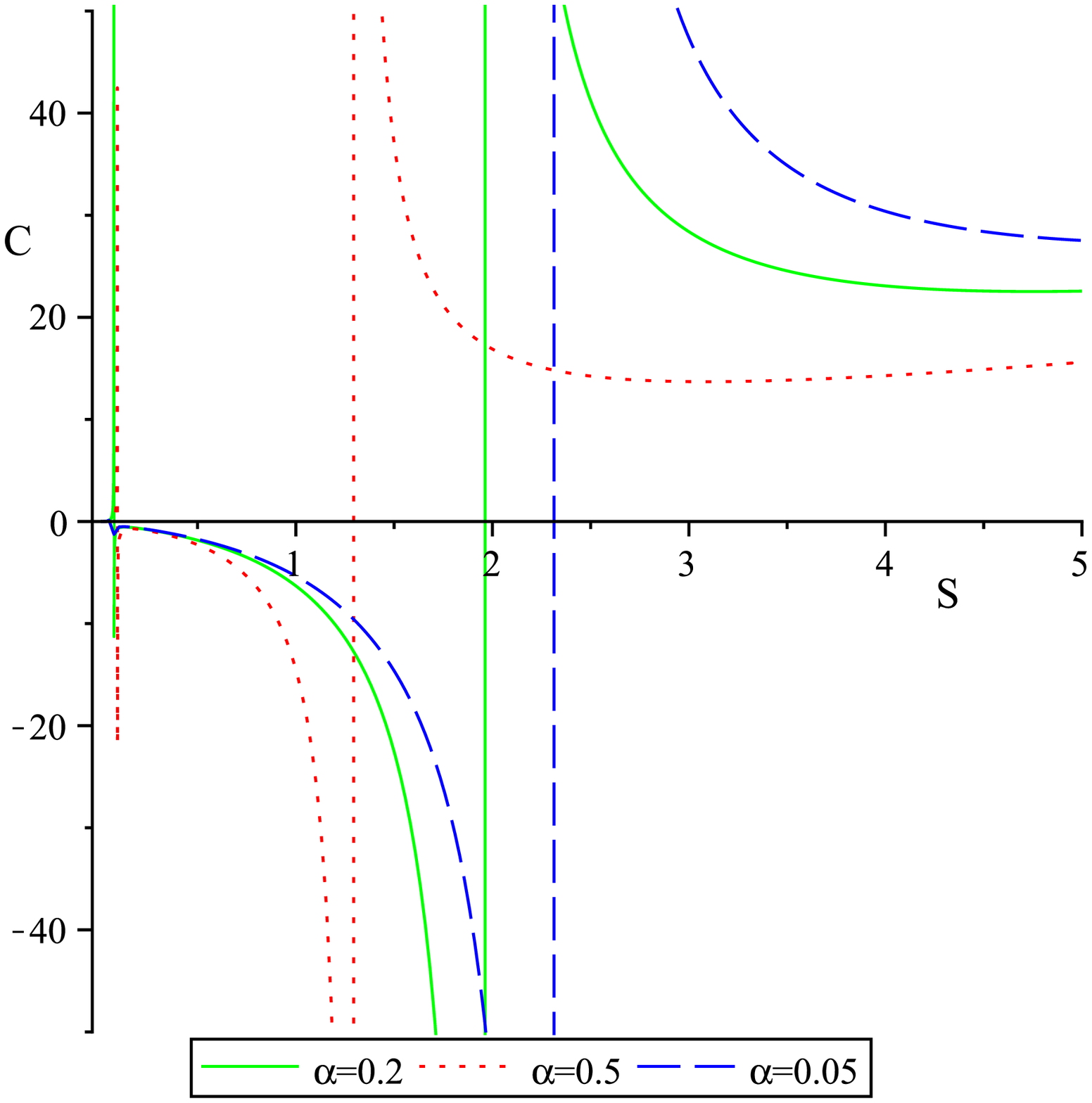,width=7cm}\caption{\small{Left plot: Heat
capacity $C$ with respect to entropy $S$ for $P=0.05$, $a=0.05$,
$\alpha=0.05$, $\omega=-\frac{2}{5}$ and different values of $Q$;
Right plot: Heat capacity $C$ with respect to entropy $S$ for
$Q=0.05$, $P=0.05$, $a=0.05$, $\omega=-\frac{2}{5}$ and different
values of $\alpha$.}}
\end{center}
\end{figure}

\section{conclusion}
In this paper, we calculated thermodynamical quantity of
Kerr-Newman-AdS black hole solution in quintessence matter. Then we
considered qualitative behavior of temperature and showed that the
rotation parameter and cosmological constant have key role on
behavior of temperature where the temperature will be always
negative for $P>0.42$ and $a>0.5$ but $\omega$ and $\alpha$
parameters have small influence. Also we investigated both types of
phase transition for different values of $\omega$ parameter. By
studying behavior of heat capacity we noticed that type one of phase
transition occurs for $P<0.42$ and $a<0.5$. We saw that the critical
point shifts to higher entropy when pressure $P$, rotation parameter
$a$ and $\alpha$ increase. Also we found that by changing parameter
$\omega$ from -1 to $-\frac{1}{3}$, the critical point shifts to
higher entropy. Then we studied type two of phase transition and
observed critical points increase by increasing parameter $\alpha$.
Also we noticed that the critical point shifts to higher entropy
when $\alpha$, $\omega$ and rotation parameter $a$ decrease.
Finally, we found that by decreasing pressure  the first critical
point shifts to lower entropy and second critical point shifts to
higher entropy.

\end{document}